# Towards Full Automation of Geometry Extraction for Biomechanical Analysis of Abdominal Aortic Aneurysm; Neural Network-Based versus Classical Methodologies

Farah Alkhatib[1], Mostafa Jamshidian[1], Donatien Le Liepvre[2], Florian Bernard[2], Ludovic Minvielle[2], Antoine Fondanèche[2], Elke R. Gizewski[3], Eva Gassner[3], Alexander Loizides[3], Maximilian Lutz[3], Florian Enzmann[4], Hozan Mufty[5, 6], Inge Fourneau[5, 6], Adam Wittek[1], Karol Miller[1,*]

[1] Intelligent Systems for Medicine Laboratory, The University of Western Australia, Perth, Western Australia, Australia

[2] Nurea, Bordeaux, France

[3] Department of Radiology, Medical University Innsbruck, Innsbruck, Austria

[4] Department of Vascular Surgery, Medical University Innsbruck, Innsbruck, Austria

[5] Department of Vascular Surgery, University Hospitals Leuven, Leuven, Belgium

[6] Department of Cardiovascular Sciences, Research Unit of Vascular Surgery, KU Leuven, Leuven, Belgium

## Abstract

Background: For the clinical adoption of stress-based rupture risk estimation in abdominal aortic aneurysms (AAAs), a fully automated pipeline, from clinical imaging to biomechanical stress computation, is essential. To this end, we investigated the impact of AI-based image segmentation methods on stress computation results in the walls of AAAs. We compared wall stress distributions and magnitudes calculated from geometry models obtained from classical semi-automated segmentation versus automated neural network-based segmentation.

Method: 16 different AAA contrast-enhanced computed tomography (CT) images were semi-automatically segmented by an analyst, taking between 15 and 40 minutes of human effort per patient, depending on image quality. The same images were automatically segmented using PRAEVAorta®2 commercial software by NUREA (https://www.nurea-soft.com/), developed based on artificial intelligence (AI) algorithms, and automatically post-processed with an in-house MATLAB code, requiring only 1-2 minutes of computer time per patient. Aneurysm wall stress calculations were automatically performed using the BioPARR software (https://bioparr.mech.uwa.edu.au/).

Results: Compared to the classical semi-automated segmentation, the automatic neural network-based segmentation leads to equivalent stress distributions, and slightly higher peak and 99th percentile maximum principal stress values. This difference is due to slightly larger lumen surface areas in automatically segmented models as compared to classical semi-automated segmentations, resulting in greater total pressure load on the wall. However, our statistical analysis indicated that the differences in AAA wall stress obtained using the two segmentation methods are not statistically significant and fall well within the typical range of inter-analyst and intra-analyst variability, justifying the use of AI-based automatic segmentation in a fully automated AAA stress computation pipeline.

Conclusions: Our findings are a steppingstone toward a fully automated pipeline for biomechanical analysis of AAAs, starting with CT scans and concluding with wall stress assessment, while at the same time highlighting the critical importance of the repeatable and accurate segmentation of the lumen, the difficult problem often underestimated by the literature.

* Corresponding author.

E-mail address: karol.miller@uwa.edu.au (K. Miller).

# 1 Introduction

Abdominal Aortic Aneurysm (AAA) disease is an asymptomatic condition, usually diagnosed accidentally through imaging indicated by some other health problem. While most AAAs never rupture or generate symptoms, those which rupture cause almost certain death (NICE (National Institute for Health and Care Excellence) 2020, Wanhainen, Van Herzeele et al. 2024). Therefore, understanding of the risk of rapid disease progression and rupture in AAAs is crucial for appropriate AAA disease management. The current management primarily relies on a maximum aortic diameter threshold, set at 55 mm for men and 50 mm for women, and its growth rate of 10 mm per year, guiding surgeons in their decision-making regarding the need for intervention (Wanhainen, Van Herzeele et al. 2024). However, the recognized limitations of this approach become evident as most AAAs with a diameter larger than 55 mm remain stable throughout the patient's lifetime (Darling, Messina et al. 1977, NICE (National Institute for Health and Care Excellence) 2020, Wanhainen, Van Herzeele et al. 2024).

This underscores the potential advantages that could arise from the use of biomechanical models to improve AAA disease progression and rupture risk assessment (Fillinger, Raghavan et al. 2002, Fillinger, Marra et al. 2003, Vande Geest, Di Martino et al. 2006, Gasser, Auer et al. 2010, Joldes, Miller et al. 2017, Miller, Mufty et al. 2020). Aortic wall stress has been the main variable of interest in almost all proposed biomechanical rupture risk indicators of AAA (Fillinger, Raghavan et al. 2002, Fillinger, Marra et al. 2003, Vande Geest, Di Martino et al. 2006, Gasser, Auer et al. 2010, Joldes, Miller et al. 2017). Computation of patient-specific aortic wall stress requires patient-specific geometry usually acquired through segmenting medical images.

The segmentation of AAA from medical images is a procedure that has historically taken a significant amount of an analyst time (Siriapisith, Kusakunniran et al. 2018). Manual segmentation of the aneurysm walls, which involves tracing the boundaries of the wall on each image slice, is a subjective process that can take very significant amount of time. Even semi-automatic segmentation methods using well-established intensity or intensity gradient-based classical algorithms require often as much as 45 minutes of work by a trained analyst per patient (Joldes, Miller et al. 2017). While it may be acceptable for research purposes, it is not practical for routine clinical use.

Artificial intelligence (AI)-based algorithms have emerged as promising tools for automatically segmenting aneurysms without user input (López-Linares, Aranjuelo et al. 2018, Lareyre, Adam et al. 2019, Caradu, Spampinato et al. 2021, Brutti, Fantazzini et al. 2022). A fully automated method for locating the aortic lumen and describing the AAA—including the identification of intraluminal thrombus (ILT) and calcifications—was introduced by (Lareyre, Adam et al. 2019). Tested on a cohort of 40 AAA patients utilizing computed tomography angiography (CTA) images, the methodology demonstrated a robust correlation with outcomes obtained from manual segmentation by an expert. A fully automated algorithm for segmenting ILT from CTA images and subsequent analysis of AAA geometry was recently proposed by (Brutti, Fantazzini et al. 2022). They implemented a deep-learning-based pipeline that uses CTA scans to localize and segment the thrombus. Eight CTA scans were used to validate this approach after it was trained on 63 CTA scans.

The accuracy and reproducibility of the segmentation are crucial for ensuring the validity of stress calculations in the AAA biomechanical analysis (Hodge, Tan et al. 2023). Our recent study (Hodge, Tan et al. 2023) revealed differences in the distribution of the calculated stress in the aneurysm wall between the segmentations done by different analysts. Notably, the differences in stress computed using geometries derived from classical and AI-based algorithm segmentations have yet to be analyzed.

In this study, we undertake a comparative aneurysm wall stress analysis of 16 AAA geometries extracted through semi-automated segmentation by an analyst and a fully automated segmentation employing the AI algorithm developed by Nurea (https://www.nurea-soft.com/). The computed aneurysm wall stress from both sets of geometries is compared using peak and 99th percentile of maximum principal stress and overall stress distributions. We used the method embedded in the freely-available open-source software platform BioPARR – Biomechanics-based Prediction of Aneurysm Rupture Risk (https://bioparr.mech.uwa.edu.au/) (Joldes, Miller et al. 2016, Joldes, Miller et al. 2017) to perform the stress calculations of the aneurysm wall.

The remainder of the paper is organized as follows: In Materials and Methods section, we describe the patient-specific AAA model generation methods, covering both the semi-automatic segmentation and the fully automatic neural network-based segmentation methods. We also discuss stress computation in the AAA wall, which includes finite element meshes, boundary conditions and load, material models and properties, and a comparison of aneurysm wall stress obtained from the models generated using the semi-automatic segmentation and the fully automatic neural network-based segmentation. In Results section, we present the patient-specific AAA geometries, patient-specific AAA finite element meshes, and aneurysm wall stress analysis for the 16 patients, using the two methods of model generation. Discussion section includes qualitative and statistical analysis of the similarities and differences between AAA geometries, AAA meshes, and lumen segmentations obtained using the semi-automatic and fully automatic neural network-based segmentation methods. The Conclusions section summarizes the key findings of the study.

# 2 Materials and Methods

## 2.1 Patient-Specific Abdominal Aortic Aneurysm (AAA) Geometries

In the study, 16 patients diagnosed with abdominal aortic aneurysm disease were included. We utilized anonymized contrast-enhanced CTA image datasets of these patients. Referring to Table 1, for patients 1-10 (previously analyzed in (Alkhatib, Wittek et al. 2023)), we used the systolic phase of 4D-CTA images provided by the Fiona Stanley Hospital (Western Australia, Australia), obtained with the Siemens SOMATOM Definition Flash CT Scanner (Siemens Healthineers AG, Forchheim, Germany). For patients 11-13, we used the systolic phase of 4D-CTA images provided by University Hospitals Leuven (Leuven, Belgium), obtained with the Siemens SOMATOM Force CT Scanner (Siemens Healthineers AG, Forchheim, Germany). For patients 14-16, we used the systolic phase of 4D-CTA images provided by Tirol Kliniken of Innsbruck Medical University (Innsbruck, Austria), obtained with the Siemens SOMATOM Drive CT Scanner (Siemens Healthineers AG, Forchheim, Germany). The maximum diameter of the aneurysm for each patient, currently serving as the main indication for surgery, is given in Table 1. Patients provided their informed consent prior to their involvement in the research. The study was conducted in accordance with the Declaration of Helsinki. The protocols were approved by Human Research Ethics and Governance at South Metropolitan Health Service (HREC-SMHS) (approval code RGS3501), Human Research Ethics Office at The University of Western Australia (approval code RA/4/20/5913), Ethics

Committee Research UZ/KU Leuven (approval code S63163), and Ethics Committee of Medical University Innsbruck (approval code 1271/2023). **Table 1** summarizes the medical image properties for each patient. The collected CTA images differed in size and resolution between individuals, contingent on the permitted radiation dose received by each patient which depends on patient's weight and height. **Table 1** also provides the maximum AAA diameter for every patient.

**Table 1** Computed tomography angiography (CTA) image dimensions and resolutions for the studied abdominal aortic aneurysm (AAA) patients; and the mean arterial blood pressure in kPa and AAA maximum diameter in mm for all patients.

| Patient no. | Image dimensions | Image spacing/ resolution (mm) | Mean blood pressure (kPa) | AAA maximum diameter (mm) |
|---|---|---|---|---|
| **1** | 512×512×1156 | 0.39×0.39×0.20 | 12 | 46.7 |
| **2** | 512×512×169 | 0.63×0.63×0.94 | 13 | 67.2 |
| **3** | 256×256×254 | 1.18×1.18×1.00 | 13 | 65.8 |
| **4** | 256×256×177 | 0.81×0.81×0.42 | 14 | 59.3 |
| **5** | 256×256×482 | 1.82×1.82×1.00 | 14 | 52.7 |
| **6** | 256×256×452 | 1.55×1.55×1.00 | 15 | 52.1 |
| **7** | 256×256×488 | 1.53×1.53×1.00 | 9 | 47.1 |
| **8** | 256×256×188 | 1.25×1.25×1.00 | 13 | 56.7 |
| **9** | 512×512×160 | 0.63×0.63×1.00 | 12 | 54.4 |
| **10** | 512×512×174 | 0.62×0.62×1.00 | 13 | 42.7 |
| **11** | 512×512×833 | 0.67×0.67×0.70 | 14 | 51.8 |
| **12** | 512×512×771 | 0.71×0.71×0.70 | 9 | 51.9 |
| **13** | 512×512×156 | 0.72×0.72×3.00 | 16 | 52.7 |
| **14** | 512×512×172 | 0.23×0.23×1.00 | 16 | 50.6 |
| **15** | 512×512×206 | 0.23×0.23×1.00 | 16 | 50.3 |
| **16** | 512×512×206 | 0.23×0.23×1.00 | 16 | 56.1 |

#### 2.1.1 Semi-automatic intensity-based segmentation

The classical semi-automated segmentation was carried out using 3D Slicer, an open-source software platform for medical images informatics, image processing, and three-dimensional visualization (https://www.slicer.org/) (Fedorov, Beichel et al. 2012). This classical segmentation approach has been employed in our previous studies (Joldes, Miller et al. 2017, Miller, Mufty et al. 2020, Alkhatib, Bourantas et al. 2023, Alkhatib, Wittek et al. 2023, Hodge, Tan et al. 2023). The CTA image was imported into 3D Slicer as DICOM (Digital Imaging and Communications in Medicine) file and cropped to designate the region of interest for the AAA, which was located distal to the renal arteries and proximal to the common iliac bifurcations.

We used the intensity threshold algorithm (Lesage, Angelini et al. 2009) available in 3D Slicer to automatically segment the lumen (**Figure 1a**). The threshold range was adjusted manually by an analyst for each patient's CT image. Any additional components detected by the threshold algorithm, aside from the aortic lumen, such as vertebrae, calcifications, and veins, were removed.

We used the "Grow from seeds" algorithm within 3D Slicer, which is an improved region-growing technique of the grow-cut algorithm described in (Zhu, Kolesov et al. 2014) to segment the aneurysm wall and the ILT (**Figure 1b**). Although it was efficient in reducing the amount of human input required, manual modifications using the paint sphere brush in 3D Slicer were still needed in some instances. Gaussian smoothing was applied to the AAA segmentations, using a standard deviation of the Gaussian kernel of 0.2. Depending on image quality, human effort required to produce required segmentation was between 15 and 40 minutes.

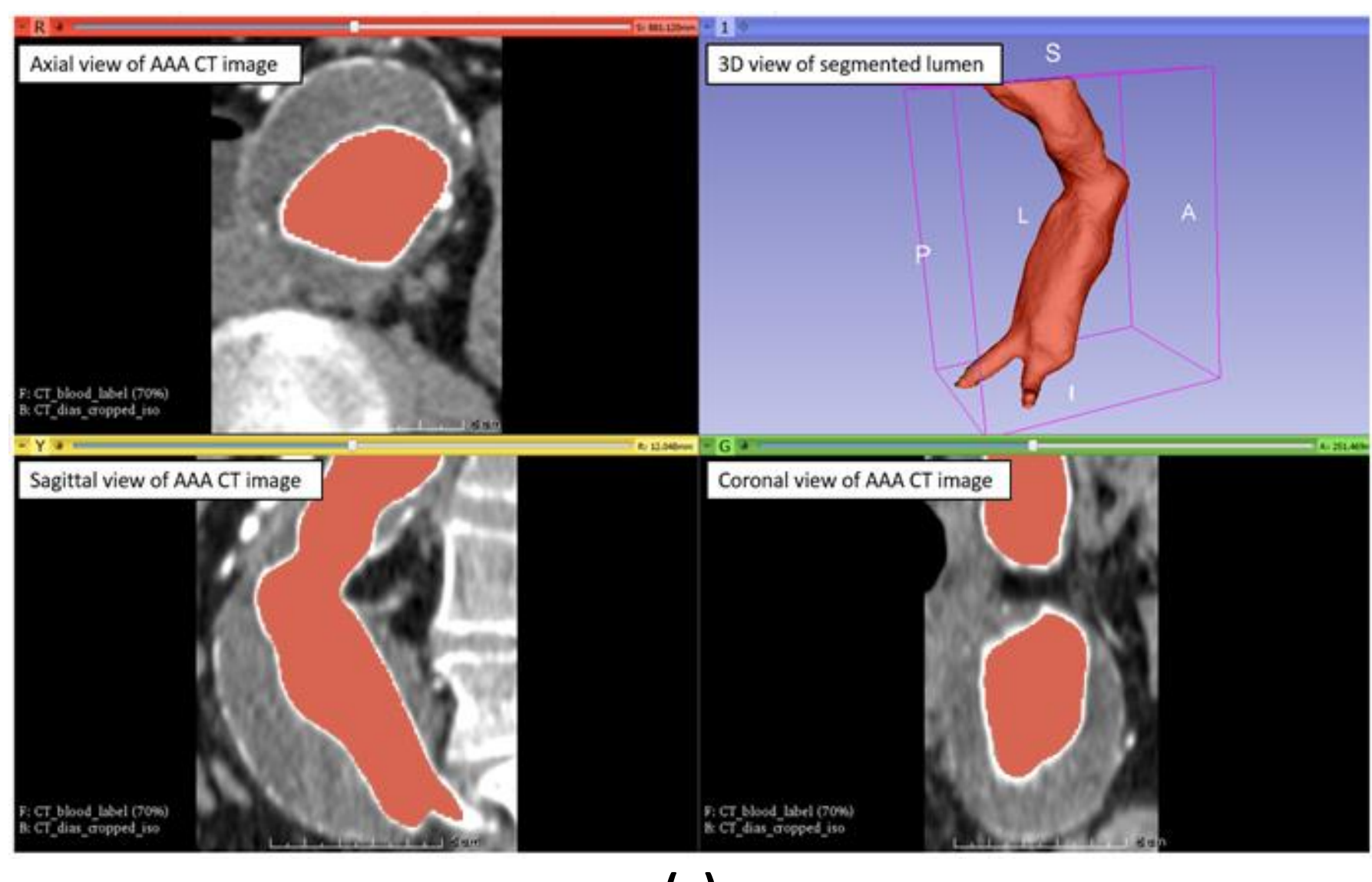

**(a)**

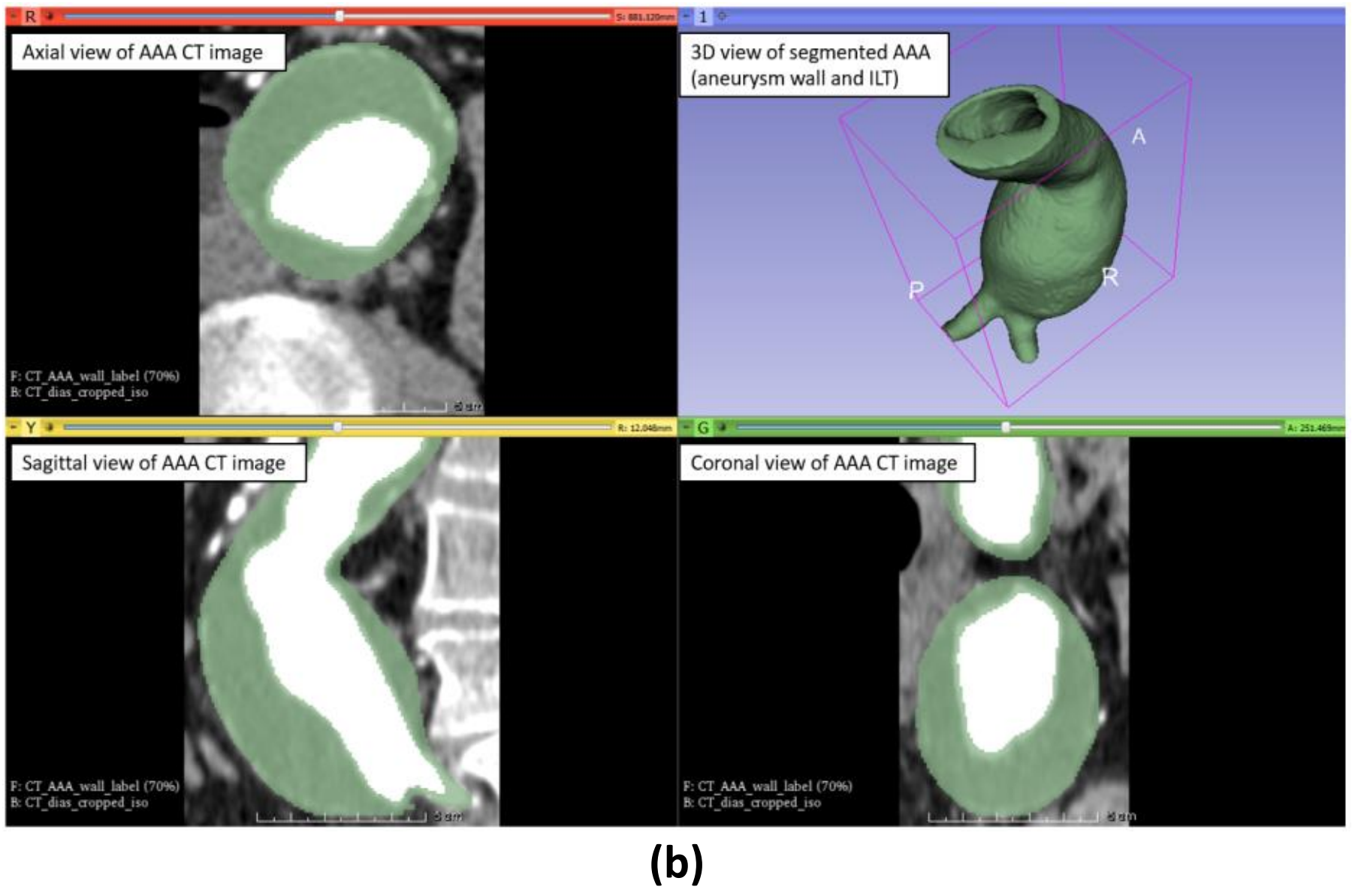

**(b)**

**Figure 1** Segmentation of abdominal aortic aneurysm (AAA) from computed tomography angiography (CTA) image using 3D Slicer software called from BioPARR (Joldes, Miller et al. 2017); **(a)** automated segmentation of the lumen using the threshold algorithm, and **(b)** semi-automated segmentation of the aneurysm wall and intraluminal thrombus (ILT).

### 2.1.2 Fully automatic neural network-based segmentation

#### *2.1.2.1 Neural network segmentation*

The fully automated segmentation was performed using PRAEVAorta®2, an automatic segmentation software developed by Nurea (https://www.nurea-soft.com/) that uses convolutional neural networks similar to the classical Unet network (Bernard and Leguay 2021, Caradu, Spampinato et al. 2021, Caradu, Pouncey et al. 2022). This software is currently commercially available. The process of this AI-based segmentation consists of five sequential steps (Caradu, Spampinato et al. 2021, Caradu, Pouncey et al. 2022): (1) image pre-processing, to filter and denoise the medical images. (2) Localization of anatomical landmarks, to define the region of interest for the AAA. (3) Lumen segmentation using simple thresholding algorithm of the voxels in the region of interest. (4) Intraluminal thrombus segmentation that uses another algorithm to identify the surrounding tissue of the lumen and discriminate the aortic wall from the vena cava and the duodenum. (5) Thresholding based on pixel intensity (>500HU) to differentiate the calcifications from the thrombus (Bernard and Leguay 2021).

This fully automated AI segmentation algorithm uses DICOM CTA images to provide three labels: (i) aneurysm wall and ILT, (ii) calcification, and (iii) lumen segmentation. **Figure 2a** shows the axial view of the segmented AAA and **Figure 2b** shows the 3D rendered geometry from the same segmented AAA.

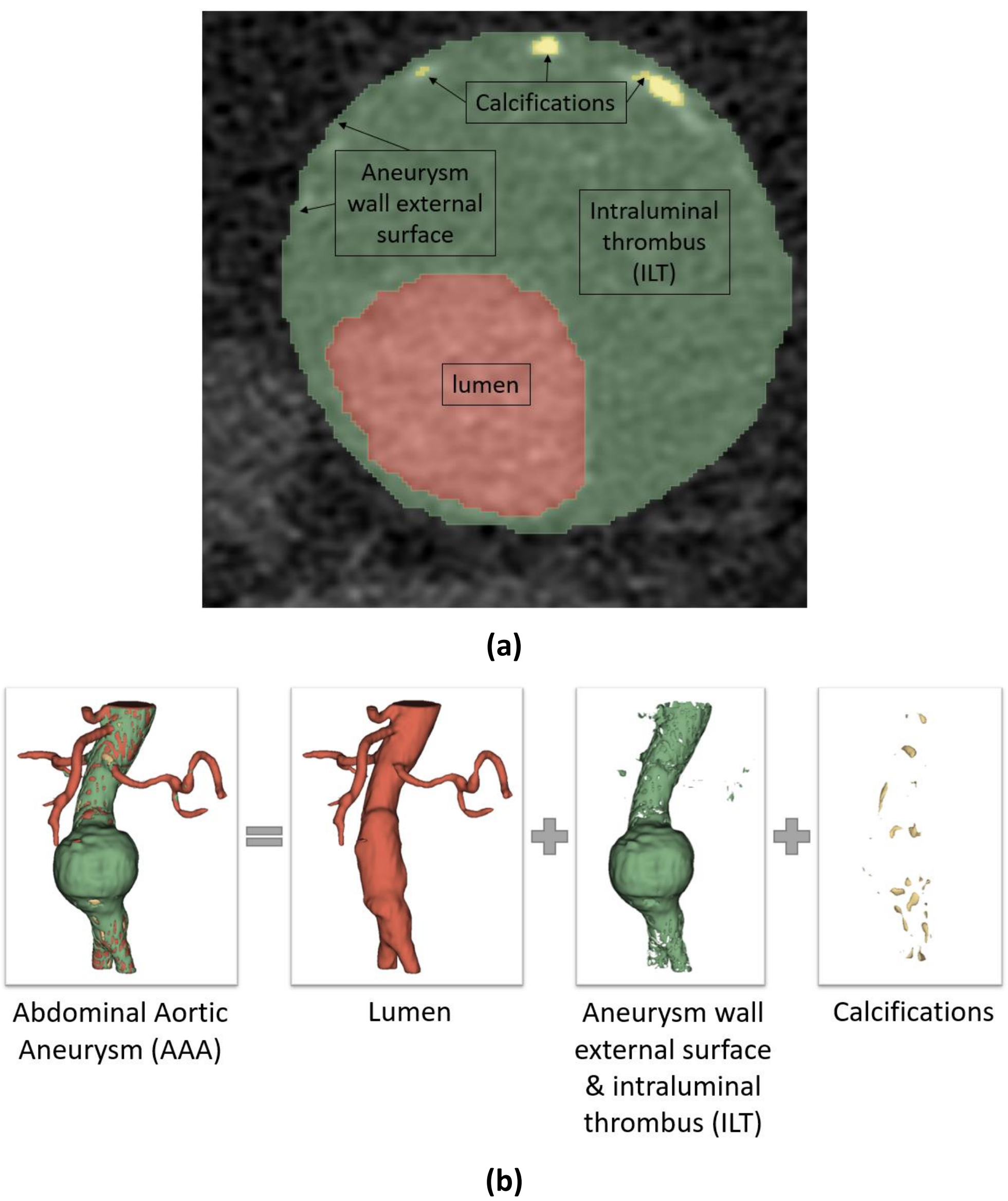

**Figure 2** The segmentation generated by the fully automated artificial intelligence (AI) segmentation algorithm (Caradu, Spampinato et al. 2021, Caradu, Pouncey et al. 2022) for the abdominal aortic aneurysm (AAA) consists of three labels: lumen is in red, aneurysm wall and intraluminal thrombus (ILT) are is in green, and calcifications are in yellow colour. **(a)** An axial view of a computed tomography angiography (CTA) scan slice showing the segmentations, and **(b)** 3D geometries of AAA rendered from the segmentations.

*2.1.2.2 Automatic post-processing of neural network segmentation*

The immediate output of the AI-based segmentation could not be directly utilized in the BioPARR automatic pipeline for stress analysis because the AI-based segmentation algorithm did not detect the aneurysm wall in certain locations (**Figure 3**), resulting in voids, holes, and significant missing parts in the aneurysm wall. Furthermore, the lumen segmentation encompassed thin layers of the aneurysm external wall, which should have been part of the aneurysm wall and ILT segmentation. Therefore, we performed automatic post-processing for the AI-based segmentations using an in-house MATLAB code developed by the Intelligent Systems for Medicine Laboratory (ISML) at The University of Western Australia (UWA). The automatic post-processing included the following main 8 steps (see **Figure 4**):

**1.** Read AI-based segmentation file and crop it using the same region of interest as used in semi-automatic segmentation. This step is needed to ensure the same aneurysm dimension in the inferior-superior direction for both semi-automated and AI-based segmentations. The AI- based segmentation contains three labels for blood (label = 1), aneurysm wall and ILT (label = 2), and calcifications (label = 3).

**2.** Create two separate binary label maps called blood and AAA. The blood binary label map is the blood label of the AI-based segmentation, and the AAA binary label map is generated by merging all three (blood, aneurysm wall and ILT, and calcifications) label maps of the AI-based segmentation.

**3.** Keep the largest island and remove smaller isolated islands from each of the blood and AAA binary label maps.

**4.** Resample both AAA and lumen binary label maps using nearest-neighbor interpolation to convert them into isotropic volumes, utilizing the smallest dimension of the original anisotropic voxel size of the image.

**5.** Blood label post-processing

**5.1.** Perform convolution smoothing of the blood label with a $N_c \times N_c \times N_c$ cubic array of ones as the convolution kernel, with $N_c = l_\mathrm{s}/l_{vox}$ (where $l_\mathrm{s} = 4.5\ mm$ and $l_\mathrm{vox}$ is the voxel size in the blood label map).

**5.2.** Perform morphological closing (filling-in holes) of the blood label using a spherical structuring element with radius $R_{vox} = r_{\mathrm{s}}/l_{vox}$ (where $r_s = 4.5\ mm$ and $l_{vox}$ is the voxel size in the blood label map).

**5.3.** Perform morphological opening (extrusion removal) of the blood label using a spherical structuring element with radius $R_{\mathrm{vox}} = r_{\mathrm{s}}/l_{\mathrm{vox}}$ (where $r_{\mathrm{s}} = 4.5\ \mathrm{mm}$ and $l_{vox}$ is the voxel size in the blood label map).

**6.** AAA label post-processing

**6.1.** Perform convolution smoothing of the AAA label with a $N_c \times N_c \times N_c$ cubic array of ones as the convolution kernel, with $N_c = l_{\mathrm{s}}/l_{vox}$ (where $l_{\mathrm{s}} = 3.0\ mm$ and $l_{\mathrm{vox}}$ is the voxel size in the AAA label map).

**6.2.** Perform morphological closing (filling-in holes) to the AAA label using a spherical structuring element with radius $R_{vox} = r_{\mathrm{s}}/l_{vox}$ (where $r_s = 4.5\ mm$ and $l_{vox}$ is the voxel size in the AAA label map).

**6.3.** Perform morphological opening (extrusion removal) of the AAA label using a spherical structuring element with radius $R_{\mathrm{vox}} = r_{\mathrm{s}}/l_{\mathrm{vox}}$ (where $r_{\mathrm{s}} = 4.5\ \mathrm{mm}$ and $l_{vox}$ is the voxel size in the AAA label map).

**7.** Final convolution smoothing of both blood label and AAA label with a $N_c \times N_c \times N_c$ cubic array of ones as the convolution kernel, with $N_c = l_{\mathrm{s}}/l_{vox}$ (where $l_{\mathrm{s}} = 3.0\ mm$ and $l_{\mathrm{vox}}$ is the voxel size in the AAA label map).

**8.** Write the post-processed AAA and Blood output labels. These labels are used as input to stress computation using BioPARR software.

The step-by-step automatic post-processing process for both lumen and AAA labels is shown in **Figure 4**. The same parameters were used for the AI-based segmentations of all 16 AAAs. For Patient 6, with a relatively large artifact extrusion, our post-processing did not remove all artifacts in the AI-based segmentation. This, however, did not result in significant differences in the stress distribution for Patient 6 in comparison to the remaining 15 patients, see **Results section**.

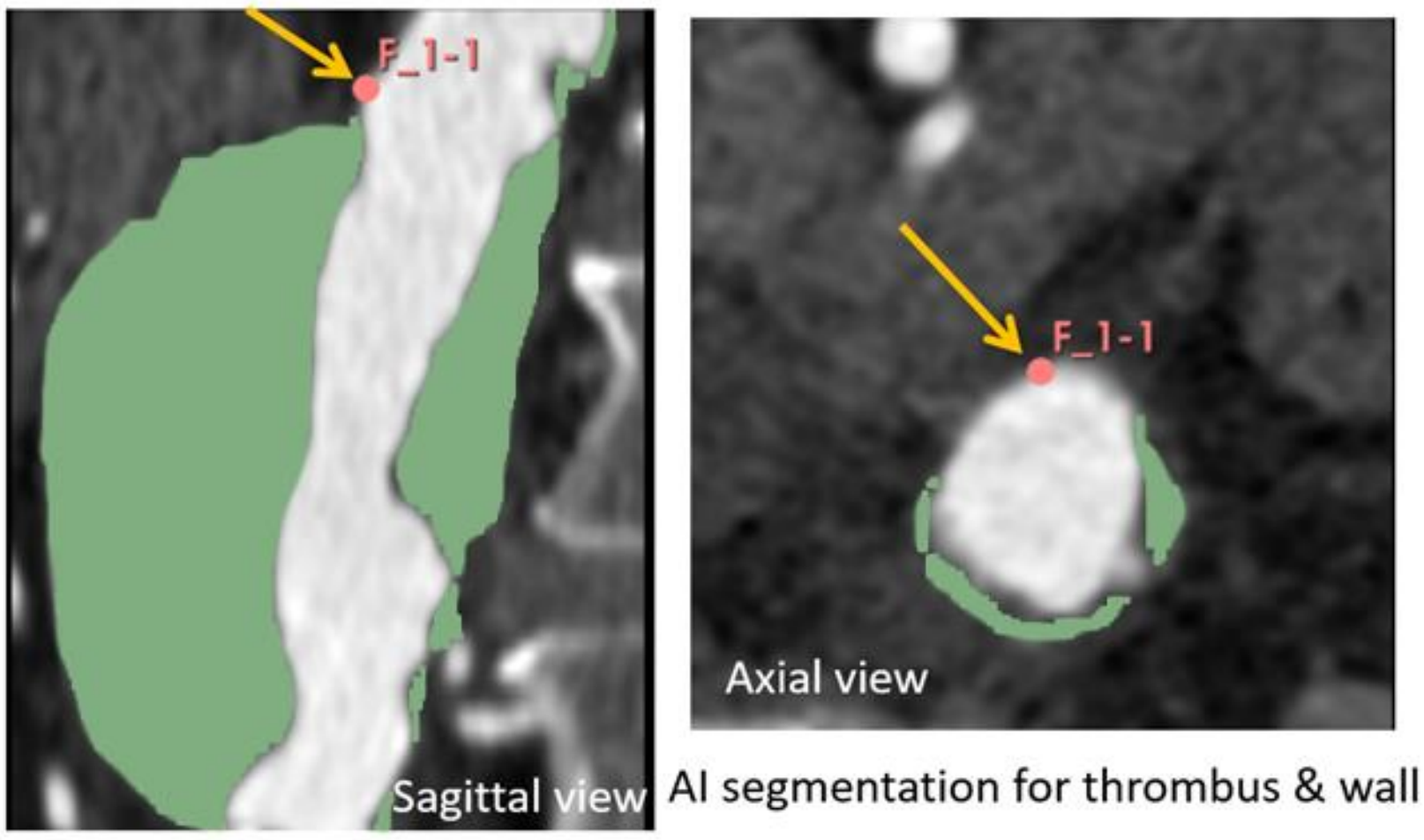

**Figure 3** The automatic segmentation by an artificial intelligence (AI) algorithm of the aneurysm wall and thrombus is in green color, and the fiducial point in both sagittal and axial views is the same point showing that the automatic algorithm could not detect the aneurysm wall in certain locations.

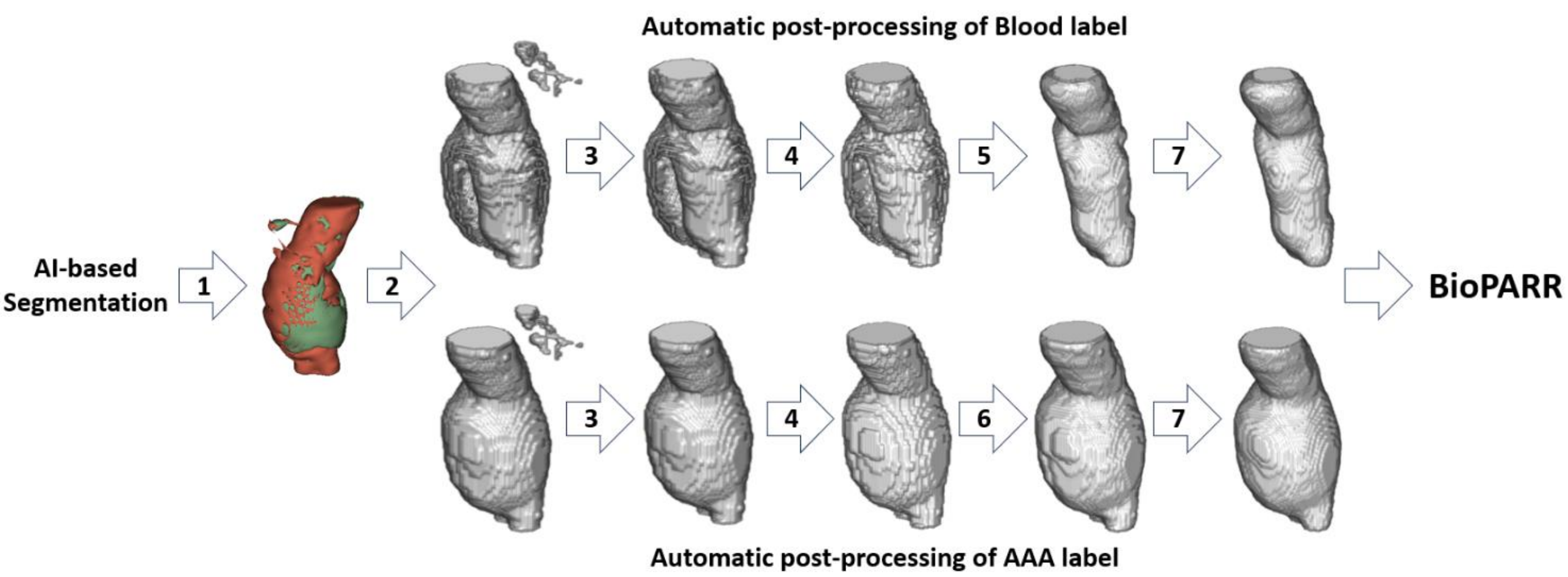

**Figure 4** Algorithmic steps for the automatic post-processing of the artificial intelligence (AI)-based segmentation.

### 2.1.3 AAA geometries

The BioPARR automatic algorithm for AAA geometry creation from image segmentation label maps, detailed in (Joldes, Miller et al. 2017), involves a three-stage process: starting with the 3D Slicer module Model Maker for label map manipulation and surface extraction (Lorensen and Cline 1998), followed by re-meshing using the surface mesh resampling software ACVDQ to refine surface quality (Valette and Chassery 2004, Valette,

Chassery et al. 2008), and finally, a custom command line interface (CLI) 3D Slicer module to generate, separate, and adjust the external and internal AAA and ILT surfaces for meshing, outputting the created AAA surfaces (**Figure 5**). A constant AAA wall thickness of 1.5 mm was assumed for the creation of all patient-specific AAA models.

Since calcifications are typically not correlated with an increased risk of AAA rupture (Maier, Gee et al. 2010), they were not taken into consideration when developing the biomechanical models in this study. This is consistent with the AAA stress computation literature (Khosla, Morris et al. 2014, Gasser 2016, Gasser, Miller et al. 2022, Wittek, Alkhatib et al. 2022).

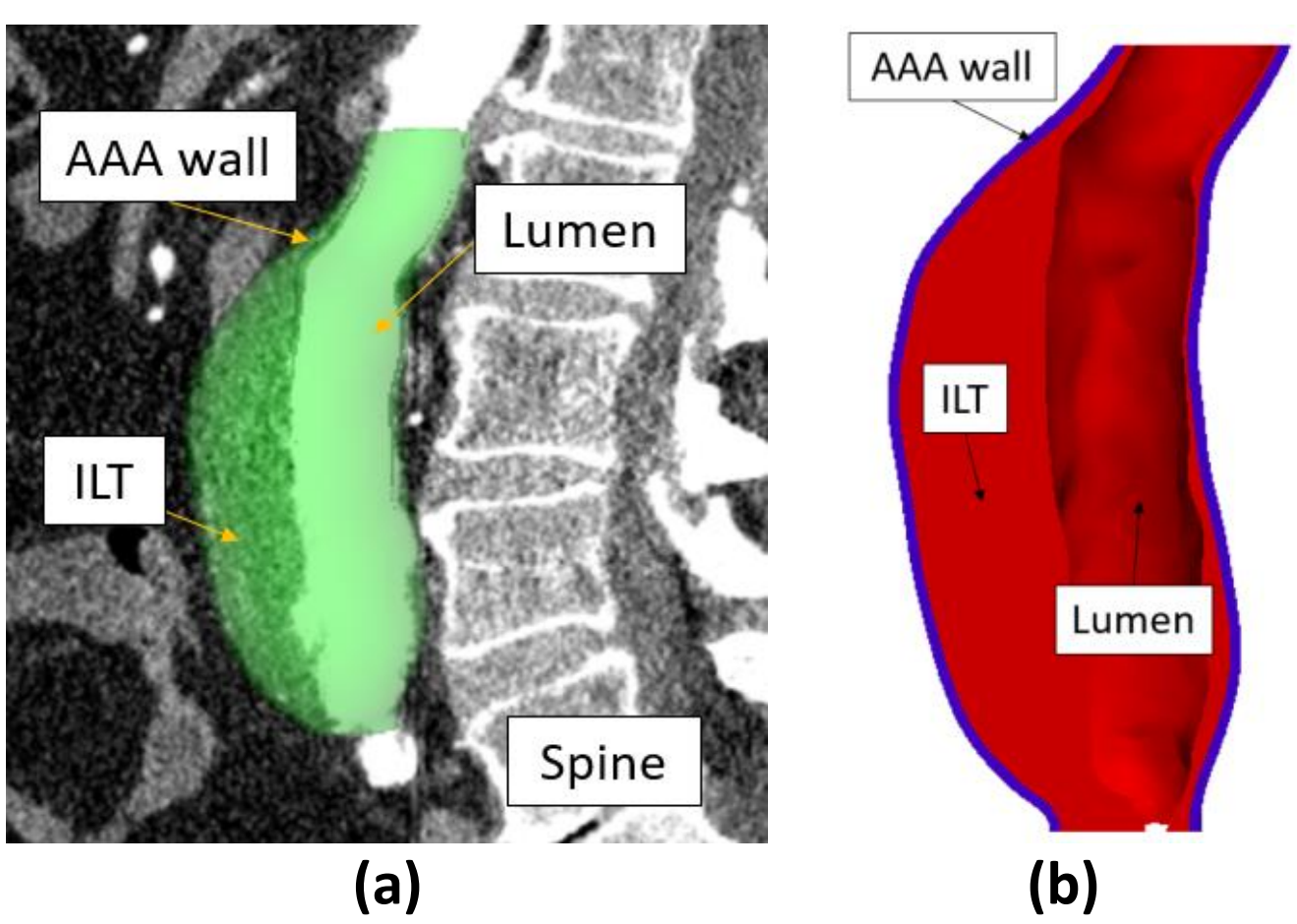

**Figure 5** Example of patient-specific abdominal aortic aneurysm (AAA) geometry (Patient 1): **(a)** AAA label map segmented from computed tomography angiography (CTA) image - sagittal view, and **(b)** a cross-section of AAA geometry showing the intraluminal thrombus (ILT), lumen, and AAA external wall (blue color) created automatically using BioPARR (Joldes, Miller et al. 2017). Figure adapted from (Alkhatib, Wittek et al. 2023).

## 2.2 Stress Computation in Abdominal Aortic Aneurysm (AAA) Wall

To calculate the stress within the aneurysm wall, we used the method embedded in the freely-available open-source software platform BioPARR – Biomechanics-based Prediction of Aneurysm Rupture Risk (https://bioparr.mech.uwa.edu.au/) (Joldes, Miller et al. 2016, Joldes, Miller et al. 2017). The methods used in BioPARR have been described in detail elsewhere (Joldes, Miller et al. 2016, Joldes, Miller et al. 2017) and used previously in our papers (Wittek, Alkhatib et al. 2022, Alkhatib, Bourantas et al. 2023, Alkhatib, Wittek et al. 2023, Hodge, Tan et al. 2023). BioPARR calls Abaqus/Standard finite element code (Simulia 2024) for a linear static finite element analysis. The aneurysm stress is extracted from the known deformed geometry (as seen on CT) and pressure load using standard linear methods that are known to be insensitive to a stress parameter in the material constitutive law (Ciarlet 1988, Lu, Zhou et al. 2007, Zelaya, Goenezen et al. 2014, Biehler, Gee et al. 2015, Joldes, Miller et al. 2016, Joldes, Miller et al. 2017, Liu, Liang et al. 2019, Wittek, Alkhatib et al. 2022).

We approximated the effects of residual stress using Fung's Uniform Stress Hypothesis (USH) (Fung 1991) by averaging the stress across the vessel wall thickness (Joldes, Noble et al. 2018). The residual stresses calculations are implemented within BioPARR platform (Joldes, Miller et al. 2016) as a simple post-processing step during the AAA biomechanical analysis.

### 2.2.1 Finite element meshes of AAA

The construction of computational grids (finite element meshes) for geometries extracted from the classical semi-automated segmentation and the automated AI-based segmentation followed the same process. Tetrahedral finite element meshes were automatically created from the AAA geometries using the open-source Gmsh mesh generator (Geuzaine and Remacle 2009, Geuzaine and Remacle 2024), called from within the BioPARR software (Joldes, Miller et al. 2017). Mesh optimization was employed during mesh generation to ensure the mesh quality. In line with our previous study (Alkhatib, Wittek et al. 2023), we determined that two quadratic tetrahedral elements through the wall thickness are sufficient for a converged solution in aneurysm wall stress calculation. For both the AAA arterial wall and the ILT, we utilized 10-noded tetrahedral elements with hybrid formulation to avoid volumetric locking (element type C3D10H in Abaqus finite element code).

### 2.2.2 Boundary conditions and load

The aneurysm's top and bottom nodes were rigidly constrained. The spine and tissues surrounding the aorta were not considered in the AAA computational biomechanics models created in this work. Such an approach is frequently used in the literature (Gasser, Auer et al. 2010, Joldes, Miller et al. 2016), and our recent study (Liddelow, Alkhatib et al. 2023) supports its applicability.

Patient-specific mean arterial blood pressure (MAP) (Cywinski 1980), measured 10 minutes before the image acquisition, was used as patient-specific load. MAP was calculated using the systolic and diastolic pressures, following the formula:

$$\text{MAP} = \frac{1}{3}\ \text{Systolic Pressure} + \frac{2}{3}\ \text{Diastolic Pressure.}$$

MAP ranged between 9 – 16 kPa with a median of 13 kPa (**Table 1**). The pressure load was uniformly distributed over the ILT luminal surface (Inzoli, Boschetti et al. 1993, Vorp, Mandarino et al. 1996).

### 2.2.3 Material models and material properties

Following previous research on arterial wall mechanics (Holzapfel, Gasser et al. 2000, Joldes, Miller et al. 2016), an almost incompressible linear elastic material model (Poisson's ratio = 0.49) for the aortic wall and ILT tissues was used. The aorta was defined as a very stiff material to ensure that the deformed geometry of AAA represented by the model remains unchanged when loaded by blood pressure (Joldes, Miller et al. 2016, Joldes, Miller et al. 2017, Miller, Mufty et al. 2020, Wittek, Alkhatib et al. 2022). The ILT was assumed to be 20 times more compliant than the AAA wall (Miller, Mufty et al. 2020).

### 2.2.4 Aneurysm wall stress comparison

In our analysis, we compared both the magnitude and distribution of the maximum principal stress calculated for AAA geometries extracted using the semi-automatic and automatic segmentations. We present the findings for the peak and 99th percentile of maximum principal stress in the aneurysm wall, as (Speelman, Bosboom et al. 2008) indicated that the 99th percentile of the aneurysm wall stress is more reproducible than the peak value of the maximum principal stress.

We generated maximum principal stress percentile plots for a quantitative assessment of aneurysm wall stress levels (Alkhatib, Wittek et al. 2023). On these plots, the horizontal axis represents the percentile rank of the maximum principal stress at a specific node within the entire population of nodes in the AAA finite element model for a given patient. The vertical axis corresponds to the actual value of the maximum principal stress at that node.

# 3 Results

## 3.1 Patient-specific Abdominal Aortic Aneurysm (AAA) Geometries

The classical semi-automated segmentation by an analyst required between 15 and 40 minutes of analyst time per patient using an off-the-shelf laptop or desktop computer. The automated segmentation was facilitated through the PRAEVAorta2® WEB portal, which simply involves uploading the images and downloading the segmentations. The AI algorithm completed the segmentation without any analyst involvement in 1 – 2 minutes of computer time per AAA image volume on Intel Xeon CPU with 8 cores, 2.5GHz clock, 32 GB of RAM, and a Tesla T4 GPU card. The subsequent automatic post-processing of the AI segmentation took less than 30 seconds per AAA image volume using an off-the-shelf laptop or desktop computer. The automatic generation of the aneurysm wall geometry, performed in 3D Slicer through BioPARR, took less than 30 seconds per AAA image volume on an off-the-shelf laptop or desktop computer. The Abaqus finite element simulation also initiated within BioPARR, took about 10 minutes per AAA on a computer with Intel(R) Core (TM) i9-12900H 2.50 GHz processor, 64 GB of RAM, and Windows 10 operating system.

**Figure 6** compares the final output of the aneurysm wall geometries from both segmentation methods. **Table 2** presents 3D rendered geometries of the segmented lumen and aneurysm wall for each patient generated through the automated AI-based algorithm. The geometries depict the results before and after post-processing. Although AI segmentations underwent a fully automated post-processing, we noticed that it is possible for the automatic AI-based segmentation to introduce artifacts such as relatively large additional segment on the outer wall of the aneurysm, as seen in Patient 6 (**Figure 6**). Such an artifact could not be corrected by the post-processing and needs to be addressed by the initial AI-based segmentation. Interestingly, the presence of this artifact did not significantly influence the computed stress distribution, see **Figure 7**.

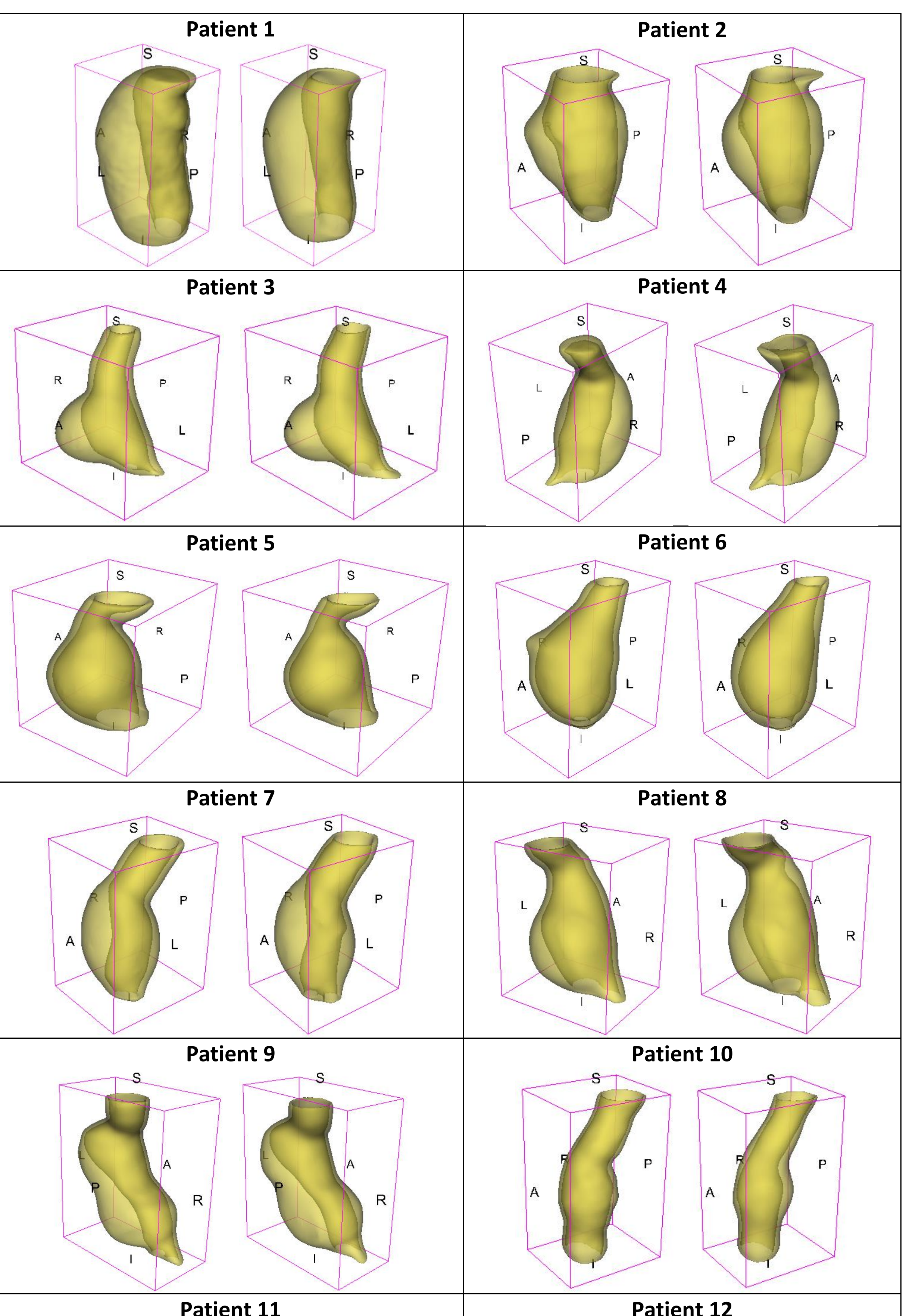

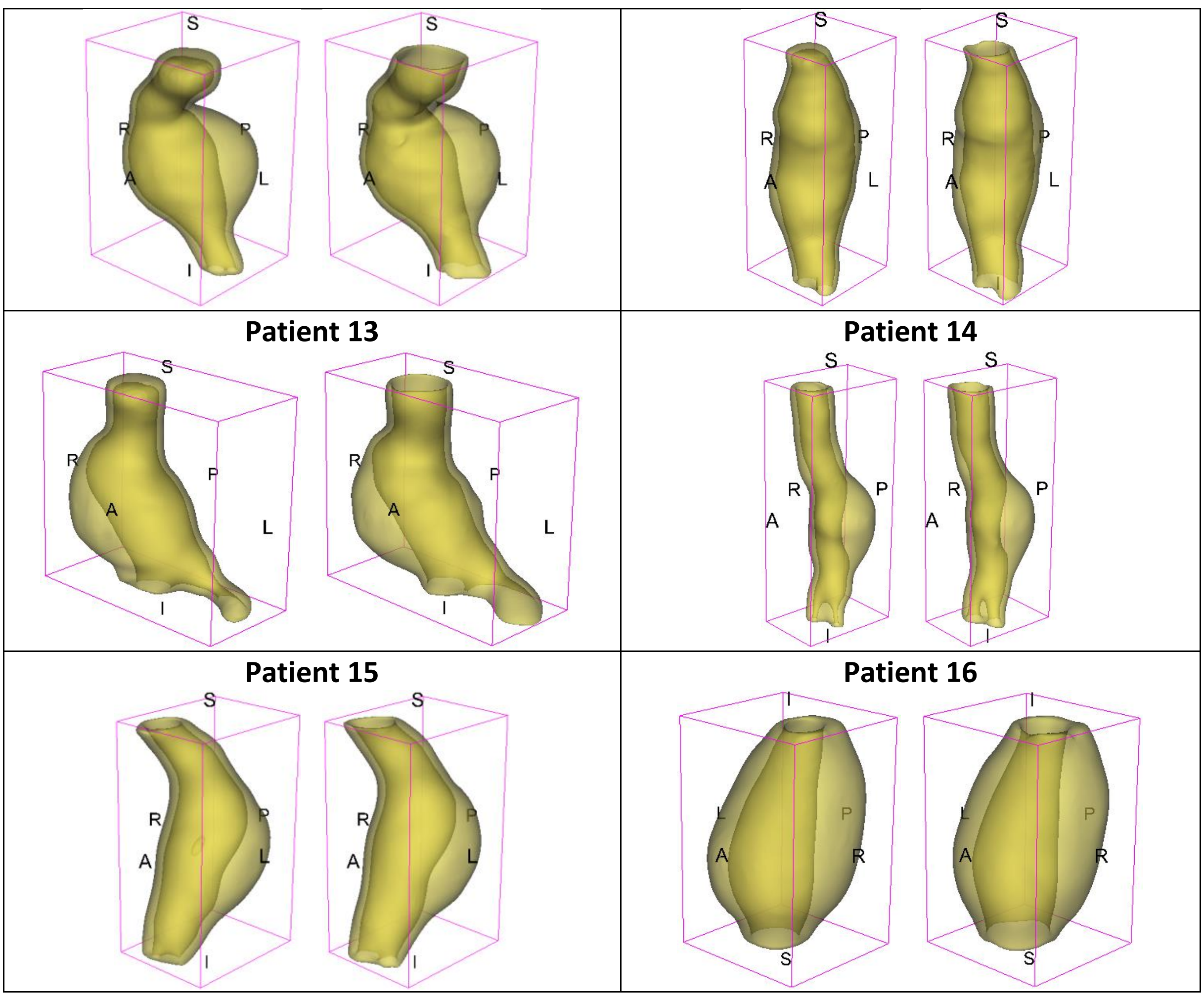

**Figure 6** Comparison of the aneurysm wall geometries created from the classical semi-automated segmentation by an analyst and the automatic segmentation using an artificial intelligence (AI) algorithm. The left-hand-side geometry is created from the automatic segmentation and the right-hand-side geometry is created from the semi-automatic segmentation.

**Table 2** 3D rendered geometries of the lumen and abdominal aortic aneurysm (AAA) wall and intraluminal thrombus (ILT) created automatically by segmenting the computed tomography angiography (CTA) images using an artificial intelligence (AI) algorithm showing the direct output from the algorithm and the output after applying a post-process step.

| **Patient no.** | **Raw segmentations created by the AI-based algorithm without any post-processing** | | **Segmentations created by the AI-based algorithm after the post-processing** | |
|---|---|---|---|---|
| | Lumen | AAA (Aneurysm wall & ILT) | Lumen | AAA (Aneurysm wall & ILT) |

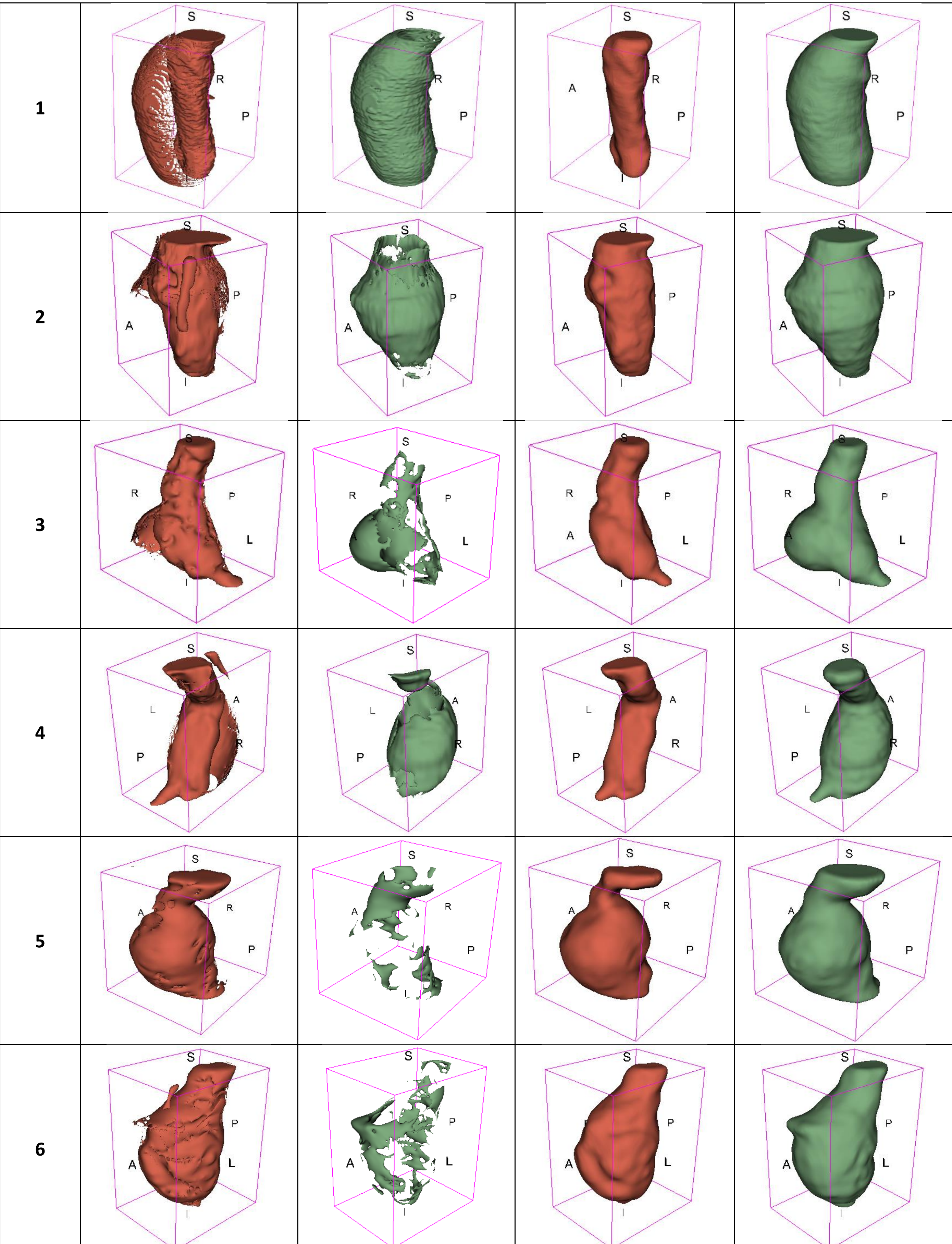

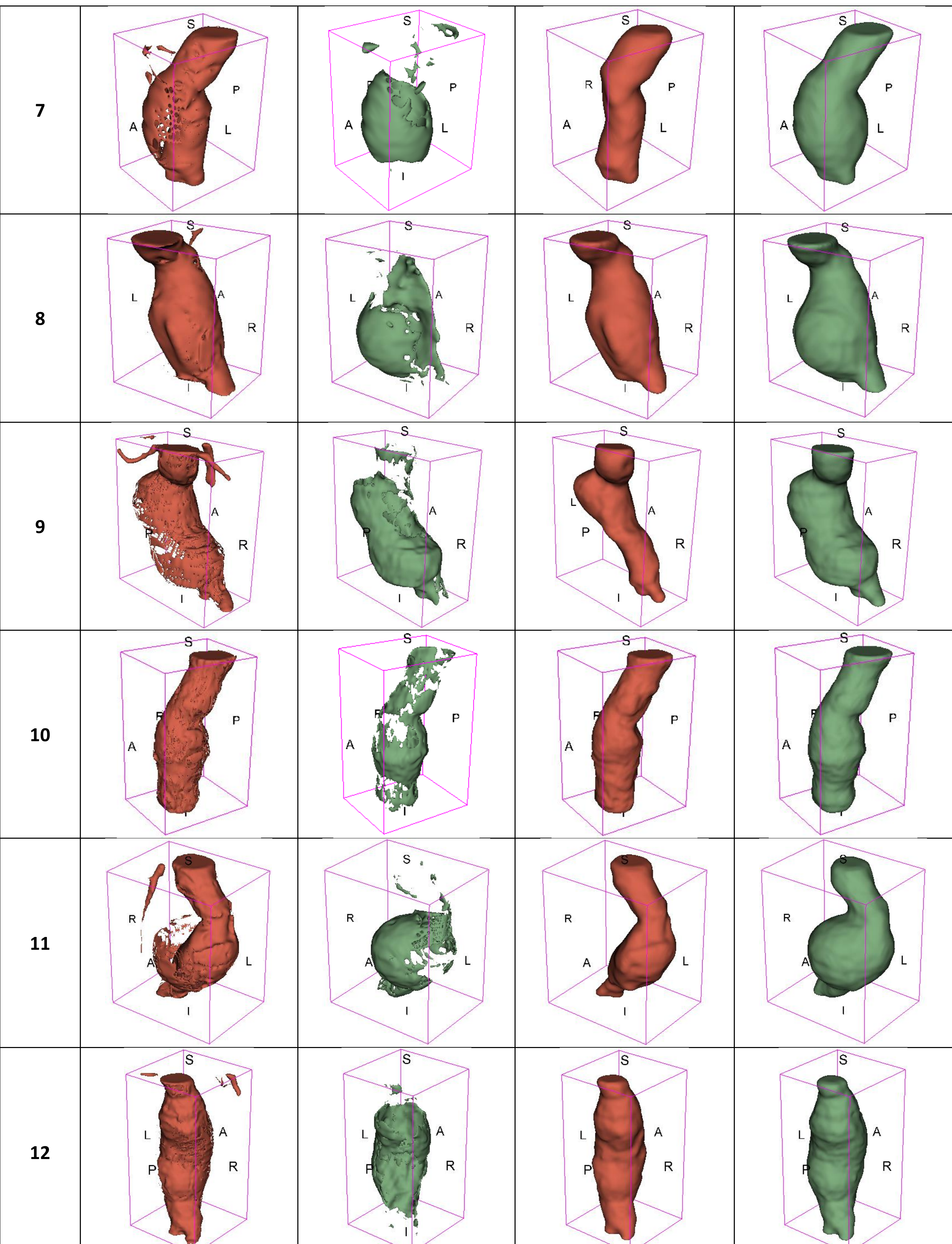

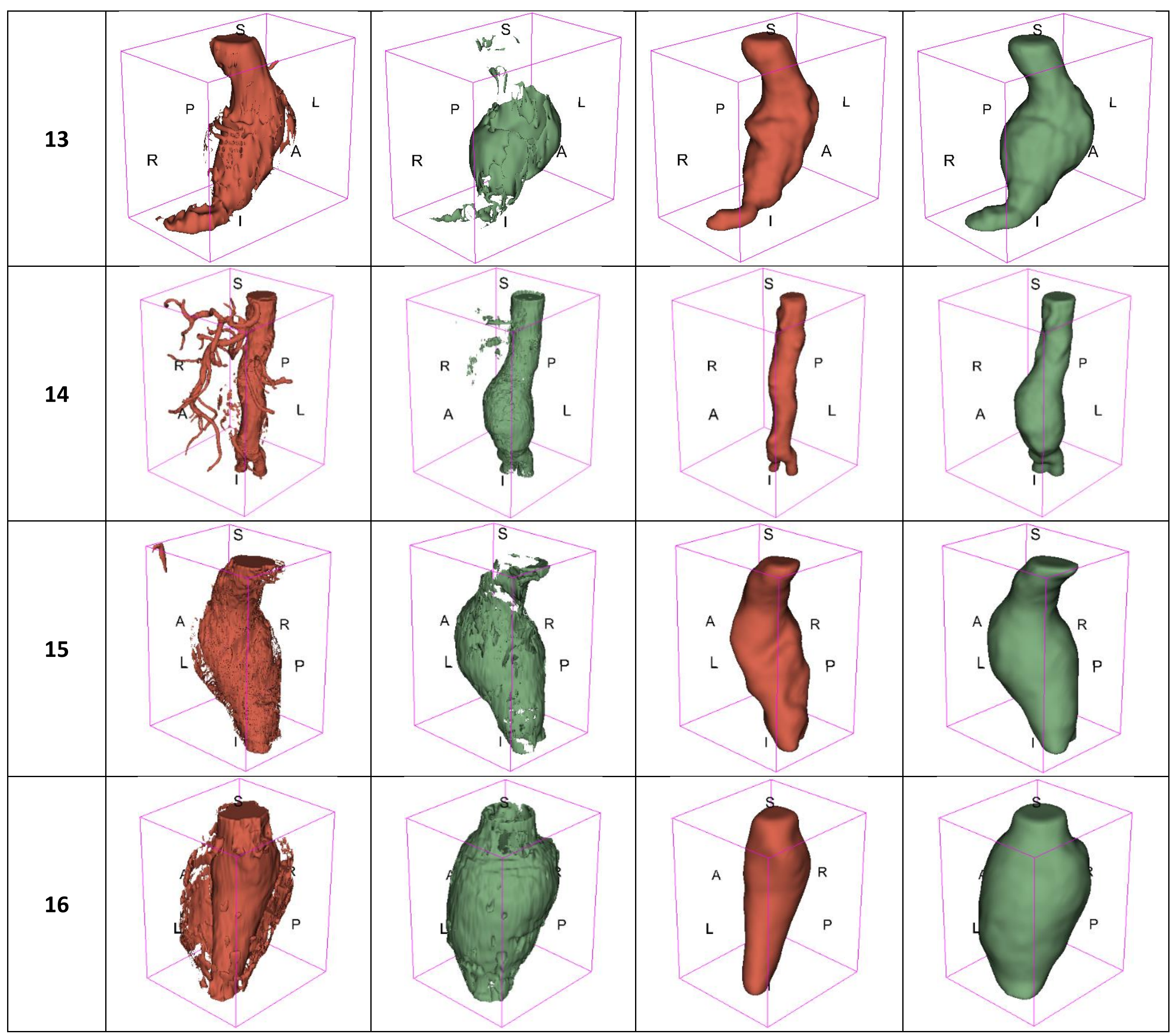

## 3.2 Finite Element Meshes of Abdominal Aortic Aneurysm (AAA)

The automated tetrahedral mesh generation with optimization (Alkhatib, Wittek et al. 2023) took between 40 – 50 minutes per patient (aneurysm wall mesh generation and ILT mesh generation) using Intel(R) Core (TM) i7-5930K CPU@3.50 GHz with 64GB of RAM running Windows 8 operating system. The number of tetrahedral elements in the aneurysm wall varied between around 520,000 – 920,000 elements depending on the patient-specific aneurysm (**Table 3**), and the number of nodes varied between 820,000 and 1,400,000. The relative difference (**Equation 1**) in the number of elements between the aneurysm walls generated by the semi-automatic segmentation and the automatic segmentation ranged between -13% and +15% (**Table 3**), while the element count for the ILT exhibits significant variability, with relative differences between -54% and +67%, particularly evident in Patient 10 (**Table 3**).

$$\text{Relative difference} = \frac{v_{auto} - v_{semi}}{v_{semi}} \times 100\% \qquad \textbf{(1)}$$

where $v_{semi}$ is the variable of interest (i.e., number of nodes or number of elements in the patient-specific finite element mesh of AAA) from the semi-automatic segmentation and $v_{auto}$ is the variable of interest from the automatic segmentation.

This difference in the total number of elements for a patient AAA model is due to differences in the surface triangulation patterns of the two models created from semi-automatic and automatic segmentations, used as input for the automatic tetrahedral (volumetric) meshing. Based on our experience, the AAA mesh refinement is sufficient to guarantee converged mesh-independent stress results (Alkhatib, Wittek et al. 2023).

**Table 3** Number of high-quality tetrahedral elements automatically constructed using Gmsh (Geuzaine and Remacle 2009, Geuzaine and Remacle 2024) called from within BioPARR (Joldes, Miller et al. 2017) for the abdominal aortic aneurysm (AAA) geometries extracted using semi-automatic segmentation and automatic artificial intelligence (AI)-based segmentation. Relative difference is calculated using **Equation 1.** ILT is intraluminal thrombus. Segmentation methods: Classical method is the semi-automated segmentation, and AI method is the automatic segmentation based on artificial intelligence (AI) algorithm.

| **Patient no.** | **Segmentation method** | **Wall** | **Relative difference (%)** | **ILT** | **Relative difference (%)** | **AAA (Wall + ILT)** | **Relative difference (%)** |
|---|---|---|---|---|---|---|---|
| **1** | Classical | 718,738 | +5.8% | 537,916 | +17.9% | 1,256,660 | +11.0% |
| | AI | 760,429 | | 634,164 | | 1,394,593 | |
| **2** | Classical | 849,479 | +4.4% | 665,981 | +28.6% | 1,515,464 | +15.0% |
| | AI | 886,653 | | 856,509 | | 1,743,162 | |
| **3** | Classical | 662,876 | -13.8% | 755,740 | -54.6% | 1,418,630 | -35.5% |
| | AI | 571,111 | | 343,292 | | 914,403 | |
| **4** | Classical | 855,299 | -3.3% | 470,965 | +16.2% | 1,326,267 | +3.6% |
| | AI | 826,671 | | 547,120 | | 1,373,791 | |
| **5** | Classical | 527,213 | +13.4% | 448,733 | -4.3% | 975,959 | +5.2% |
| | AI | 597,600 | | 429,309 | | 1,026,909 | |
| **6** | Classical | 624,894 | -0.8% | 433,014 | -3.5% | 1,057,909 | -1.9% |
| | AI | 619,726 | | 417,873 | | 1,037,599 | |
| **7** | Classical | 656,219 | -10.6% | 524,245 | -31.6% | 1,180,475 | -19.9% |
| | AI | 586,835 | | 358,575 | | 945,410 | |
| **8** | Classical | 863,937 | -11.4% | 810,900 | -36.0% | 1,674,848 | -23.3% |
| | AI | 765,167 | | 518,620 | | 1,283,787 | |
| **9** | Classical | 745,591 | -1.7% | 690,722 | +13.7% | 1,436,315 | +5.7% |
| | AI | 733,280 | | 785,059 | | 1,518,339 | |
| **10** | Classical | 796,538 | +14.9% | 658,838 | +67.3% | 1,455,391 | +38.6% |
| | AI | 915,373 | | 1,102,296 | | 2,017,669 | |
| **11** | Classical | 649,397 | -18.0% | 720,692 | -23.4% | 1,370,089 | -20.9% |
| | AI | 532,250 | | 551,962 | | 1,084,212 | |
| **12** | Classical | 962,719 | +0.6% | 827,483 | +21.8% | 1,790,202 | +10.4% |
| | AI | 968,693 | | 100,8119 | | 1,976,812 | |
| **13** | Classical | 683,489 | -7.5% | 612,385 | -12.4% | 1,295,874 | -9.8% |
| | AI | 632,214 | | 536,246 | | 1,168,460 | |
| **14** | Classical | 895,548 | -3.5% | 508,602 | -9.2% | 1,404,150 | -5.6% |
| | AI | 863,833 | | 462,060 | | 1,325,893 | |
| **15** | Classical | 610,903 | -4.7% | 362,520 | -5.8% | 973,423 | -5.1% |
| | AI | 582,239 | | 341,352 | | 923,591 | |
| **16** | Classical | 586,550 | -7.8% | 333,846 | -16.9% | 920,396 | -11.1% |
| | AI | 540,604 | | 277,284 | | 817,888 | |

### 3.3 Aneurysm Wall Stress Analysis

Here we compare the peak values, 99th percentile and distribution of the maximum principal stress in the AAA wall obtained from models developed using semi-automated and automated AI-based segmentations. The comparison was conducted for all 16 patients analyzed in this study.

The peak value of maximum principal stress (**Table 4**) was higher for 12 aneurysm walls extracted using the automatic segmentation, however 4 patients showed higher peak of maximum principal stress in the semi-automatically segmented aneurysm wall compared to the automatically segmented walls. The 99th percentile maximum principal stress followed the same behavior as the peak maximum principal stress (**Table 4**). Contour plots of the maximum principal stress for the analyzed patients are shown in **Figure 7a.** Distributions of the percentile plots of the maximum principal stress in the aneurysm walls computed using the geometries extracted by the semi-automatic and automatic segmentations were similar for all 16 patients (**Figure 7b**).

For all patients, except patient 10, the absolute relative difference in the 99th percentile maximum principal stress between the two segmentation methods is less than 15%, with an average of 7.2%. This strongly suggests that fully automatic segmentations can serve as a reliable input for AAA stress analysis pipelines compatible with existing clinical workflows.

To provide a patient-by-patient visualization of the stress results, **Figure 8** illustrates a comparison of the 99th percentile maximum principal stress, obtained from AAA geometries segmented using the two segmentation methods, for individual patients. Using the methodology behind the test for the difference in proportions (Connor 1987) our data shows that the probability of observing at least 87.5% of patients with a 99th percentile maximum principal stress difference of less than 15% between the two methods (with the true population proportion assumed to be 80%) is 81.78%.

**Table 4** Comparison of peak and 99th percentile maximum principal stress, along with relative differences (calculated using **Equation 1**), between finite element models derived from semi-automated and automatic segmentations of abdominal aortic aneurysms (AAA). Segmentation methods: Classical method is the semi-automated segmentation, and AI method is the automatic segmentation based on artificial intelligence (AI) algorithm.

| **Patient no.** | **Segmentation method** | **Peak of maximum principal stress (MPa)** | | **99th percentile maximum principal stress (MPa)** | |
|---|---|---|---|---|---|
| | | **Value** | **Relative difference (%)** | **Value** | **Relative difference (%)** |
| **1** | Classical | 0.1574 | +11.0 | 0.1486 | +7.8 |
| | AI | 0.1747 | | 0.1602 | |
| **2** | Classical | 0.2237 | +22.8 | 0.2117 | +15.0 |
| | AI | 0.2748 | | 0.2435 | |
| **3** | Classical | 0.1888 | -5.5 | 0.1701 | -1.4 |
| | AI | 0.1785 | | 0.1678 | |
| **4** | Classical | 0.2141 | +17.0 | 0.1959 | +14.0 |
| | AI | 0.2505 | | 0.2233 | |
| **5** | Classical | 0.2265 | +7.6 | 0.2073 | +5.8 |
| | AI | 0.2438 | | 0.2194 | |
| **6** | Classical | 0.2229 | +18.8 | 0.2141 | +14.3 |
| | AI | 0.2648 | | 0.2448 | |
| **7** | Classical | 0.1154 | +10.1 | 0.1089 | +5.6 |
| | AI | 0.127 | | 0.1150 | |
| **8** | Classical | 0.2145 | +3.6 | 0.2013 | +2.1 |

| | AI | 0.2223 | | 0.2056 | |
|---|---|---|---|---|---|
| **9** | Classical | 0.2087 | -9.5 | 0.1913 | -12.6 |
| | AI | 0.1888 | | 0.1672 | |
| **10** | Classical | 0.1940 | +37.0 | 0.1716 | +28.3 |
| | AI | 0.2658 | | 0.2201 | |
| **11** | Classical | 0.2098 | -19.4 | 0.1799 | -11.8 |
| | AI | 0.1691 | | 0.1587 | |
| **12** | Classical | 0.1707 | +2.9 | 0.1613 | +1.3 |
| | AI | 0.1756 | | 0.1633 | |
| **13** | Classical | 0.2506 | +11.7 | 0.2152 | +6.6 |
| | AI | 0.2799 | | 0.2293 | |
| **14** | Classical | 0.1880 | +2.8 | 0.1724 | +2.7 |
| | AI | 0.1932 | | 0.1770 | |
| **15** | Classical | 0.2340 | -1.3 | 0.2143 | -0.6 |
| | AI | 0.2310 | | 0.2131 | |
| **16** | Classical | 0.1545 | +14.7 | 0.1460 | +6.1 |
| | AI | 0.1772 | | 0.1548 | |

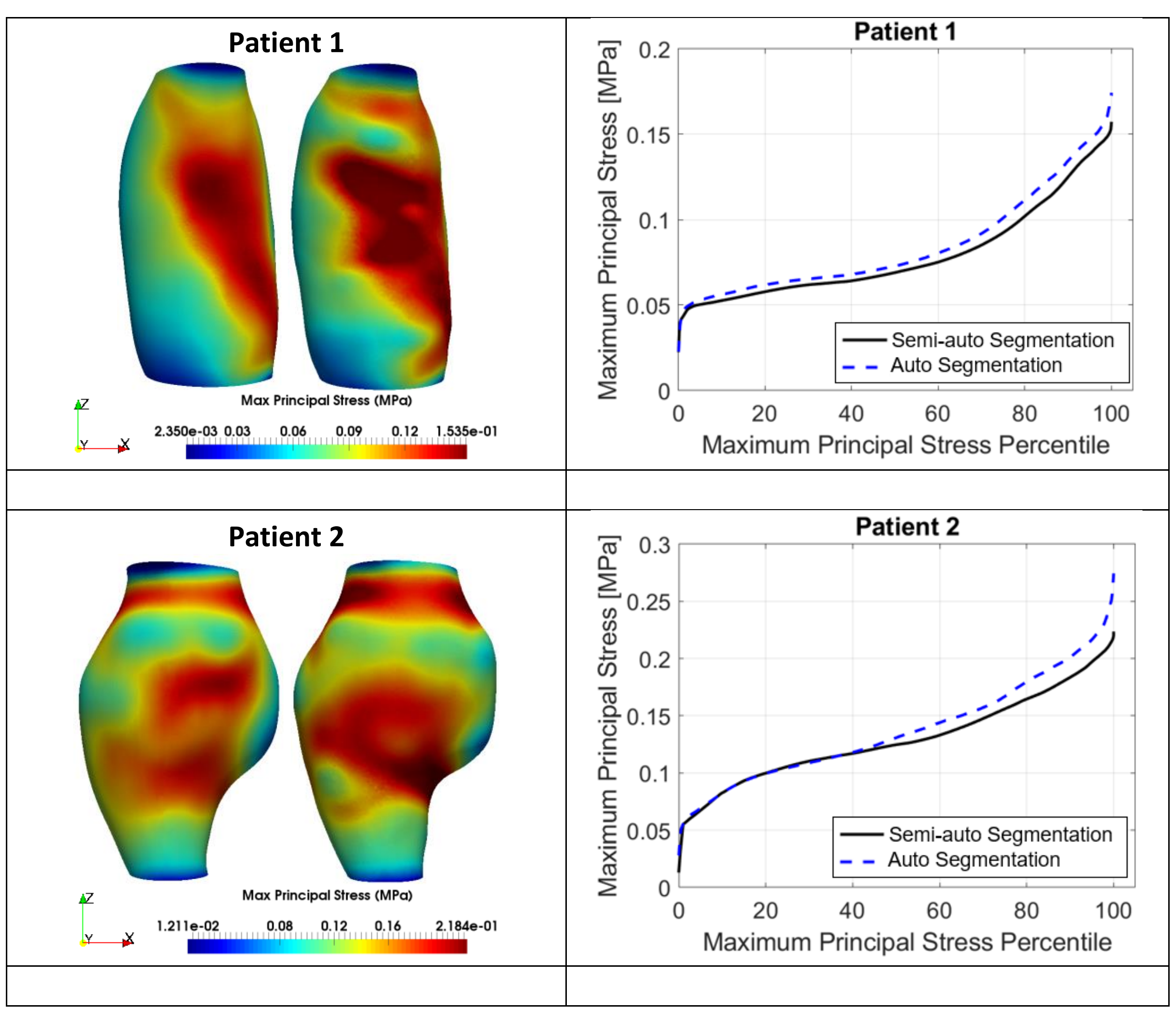

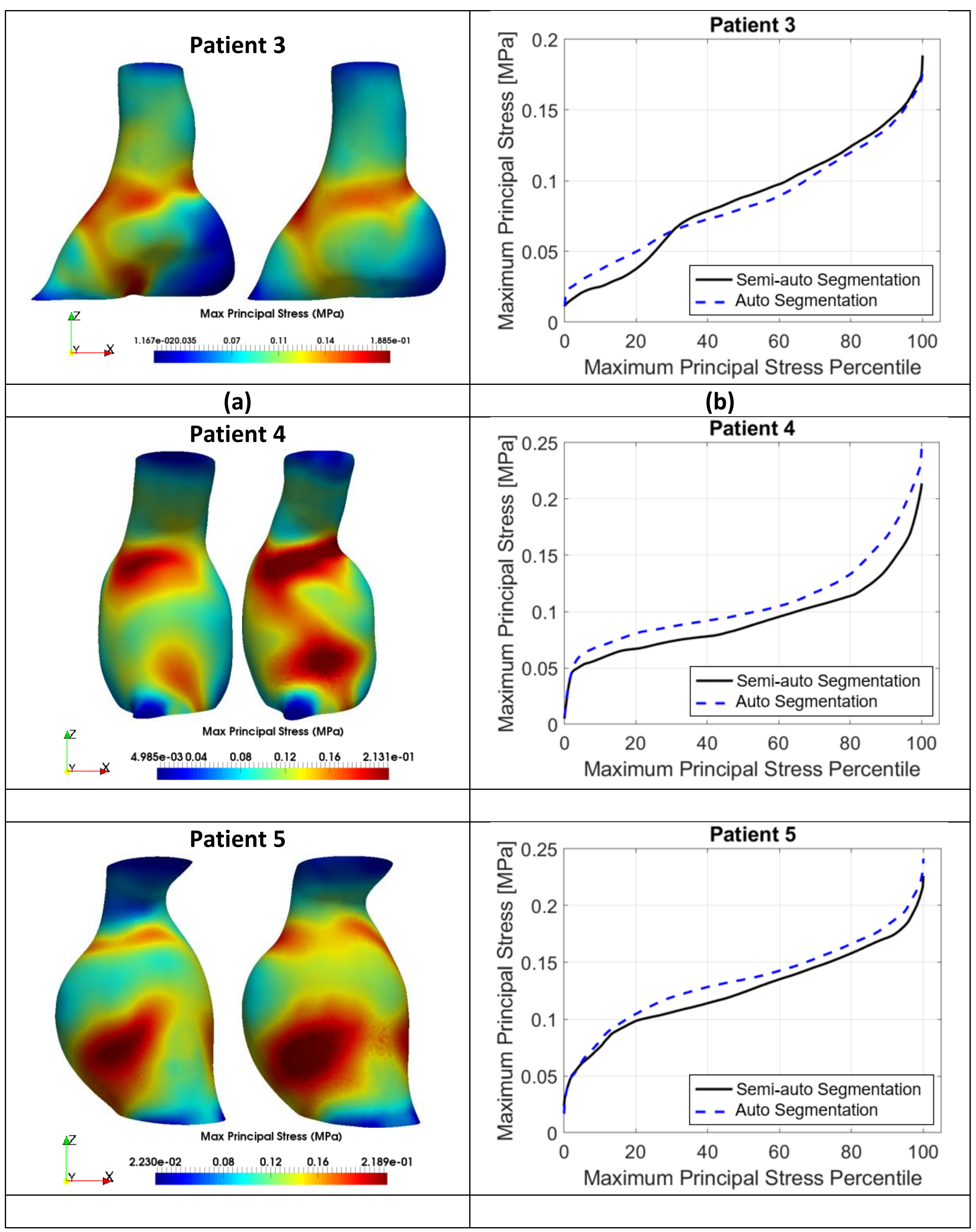

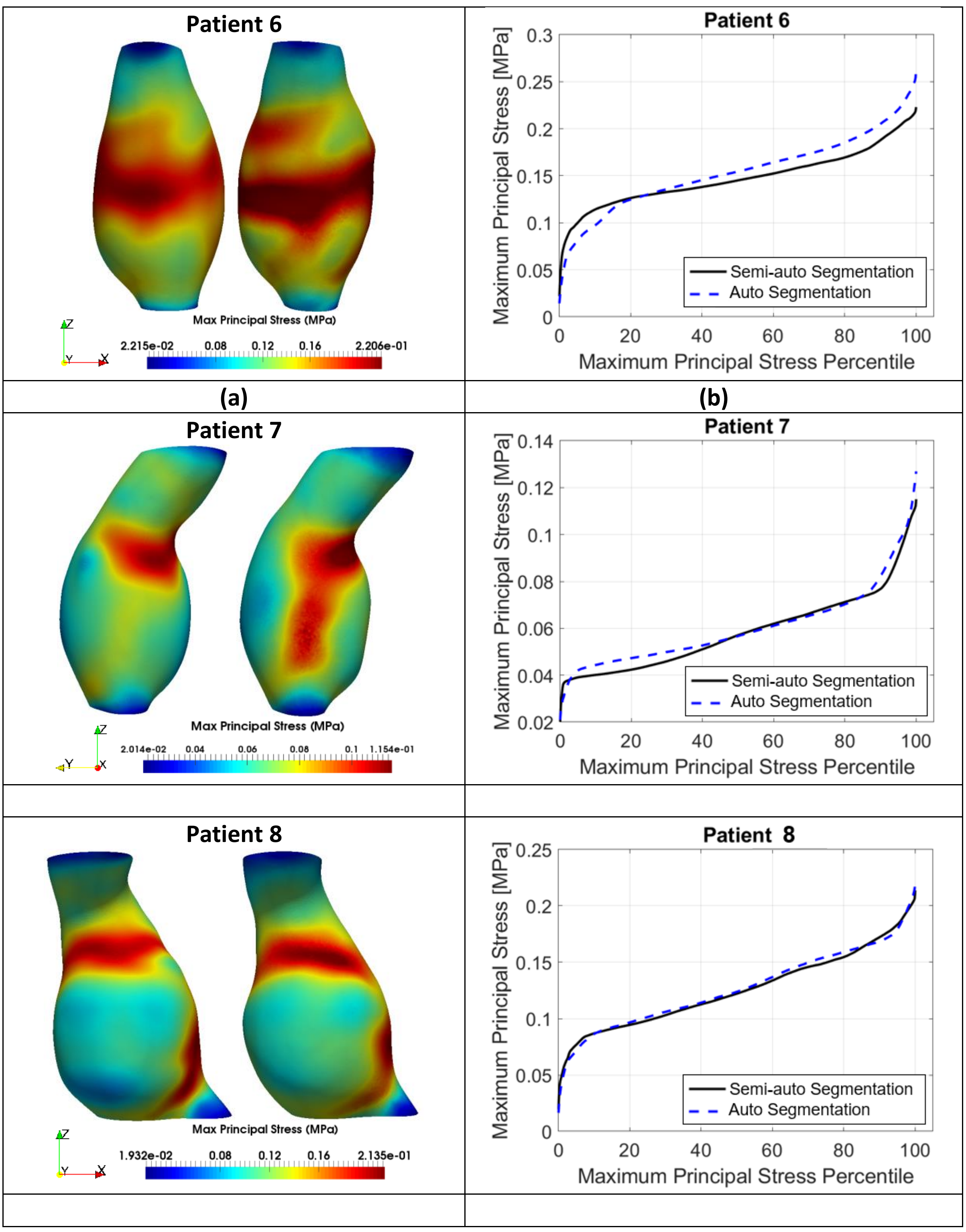

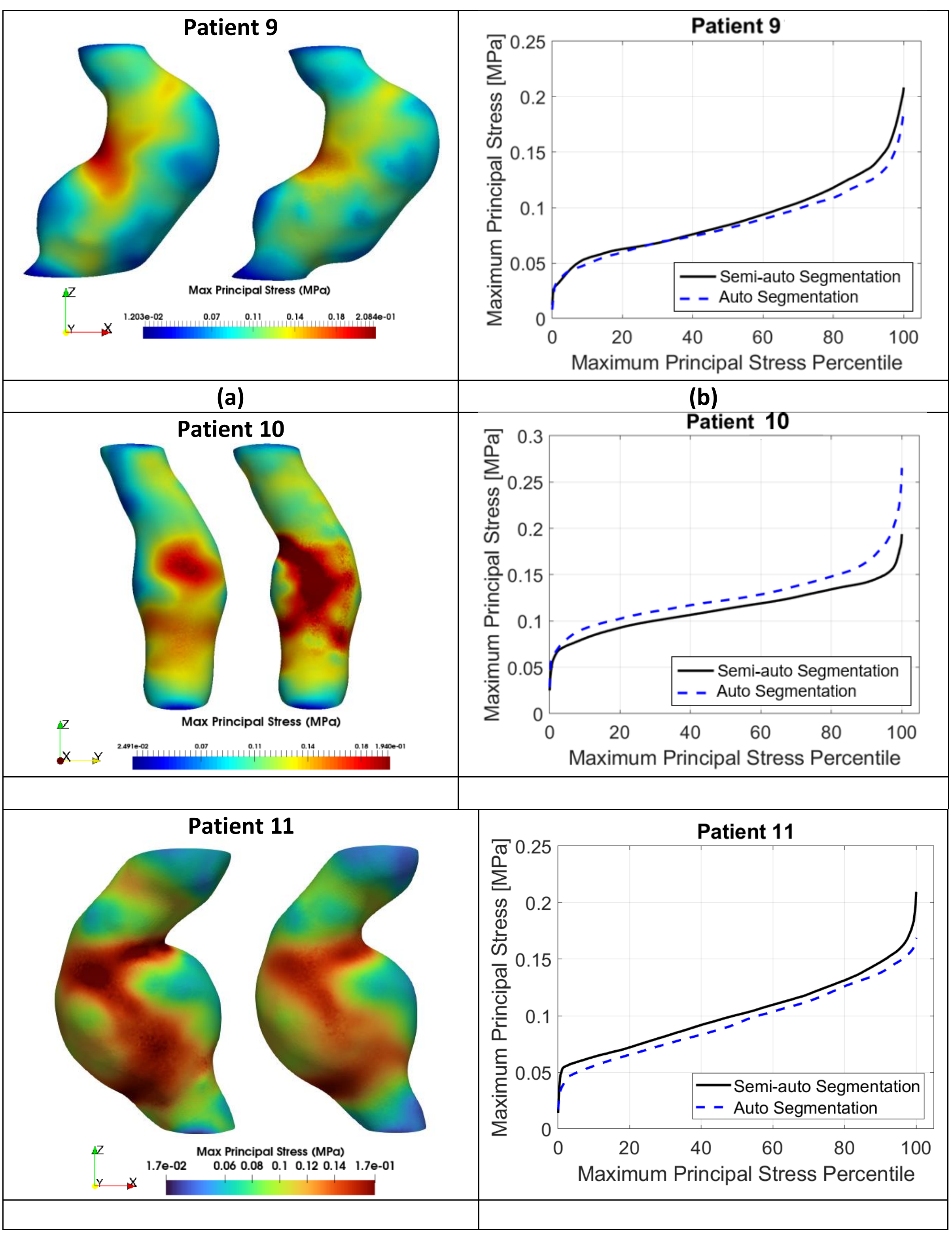

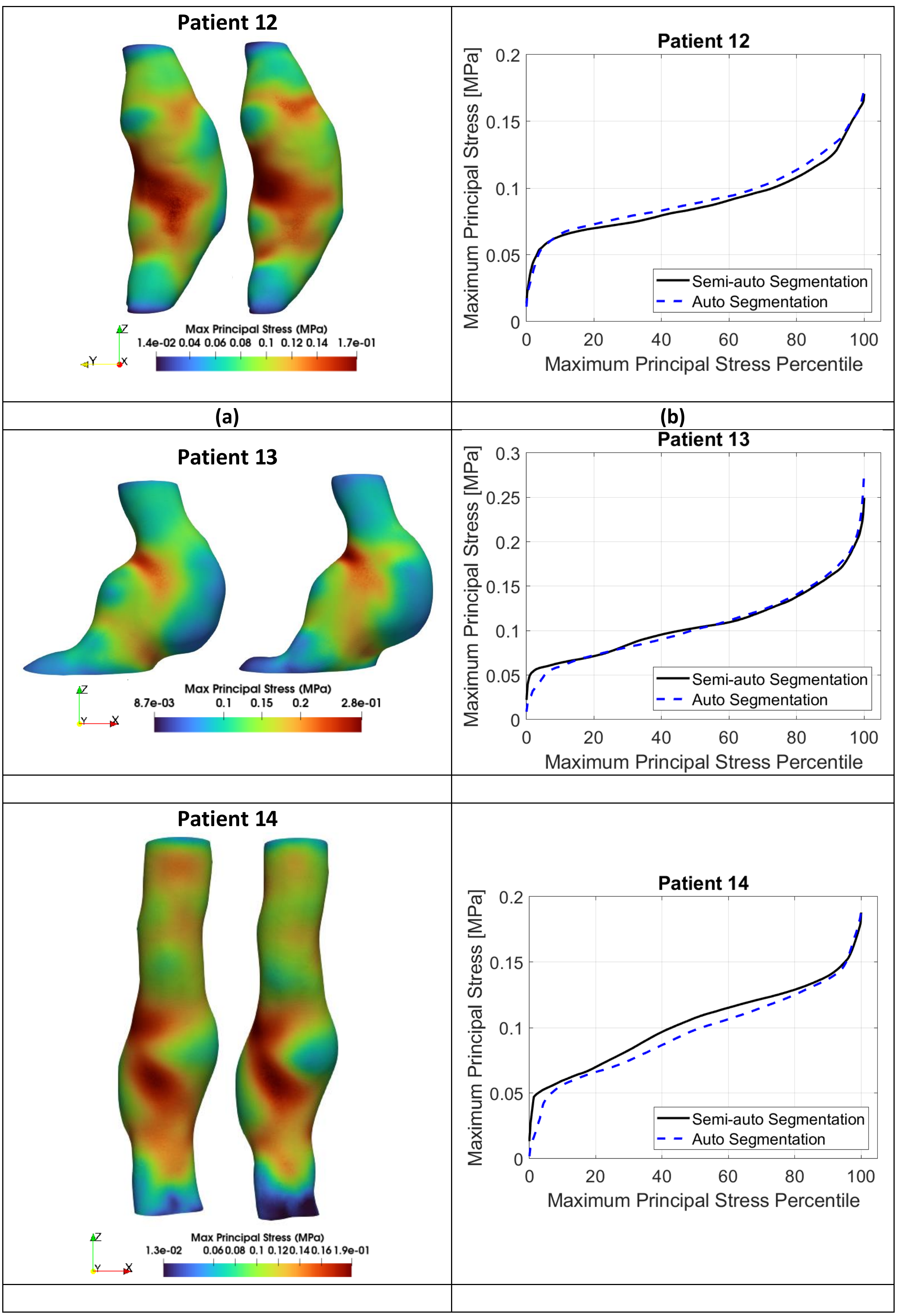

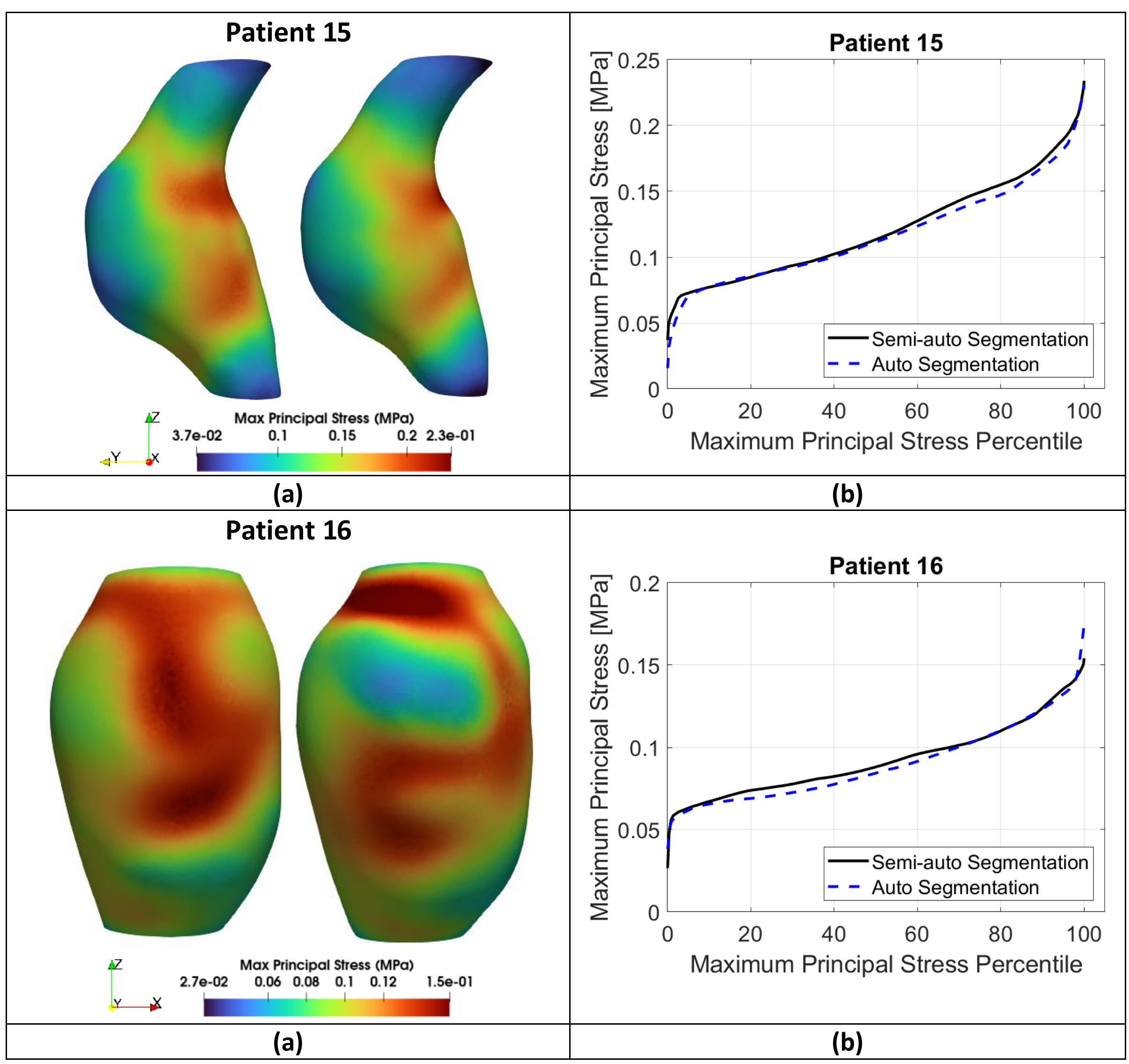

**Figure 7** Patient-specific finite element models of abdominal aortic aneurysm (AAA). **(a)** Contour plots of the maximum principal stress including residual stress in the aneurysm wall (Joldes, Noble et al. 2018) obtained using AAA geometries segmented semi-automatically (left-hand-side figures) and AAA geometries segmented automatically using artificial intelligence (AI) algorithm (right-hand-side figures); and **(b)** percentile plots of maximum principal stress for the studied AAA patients.

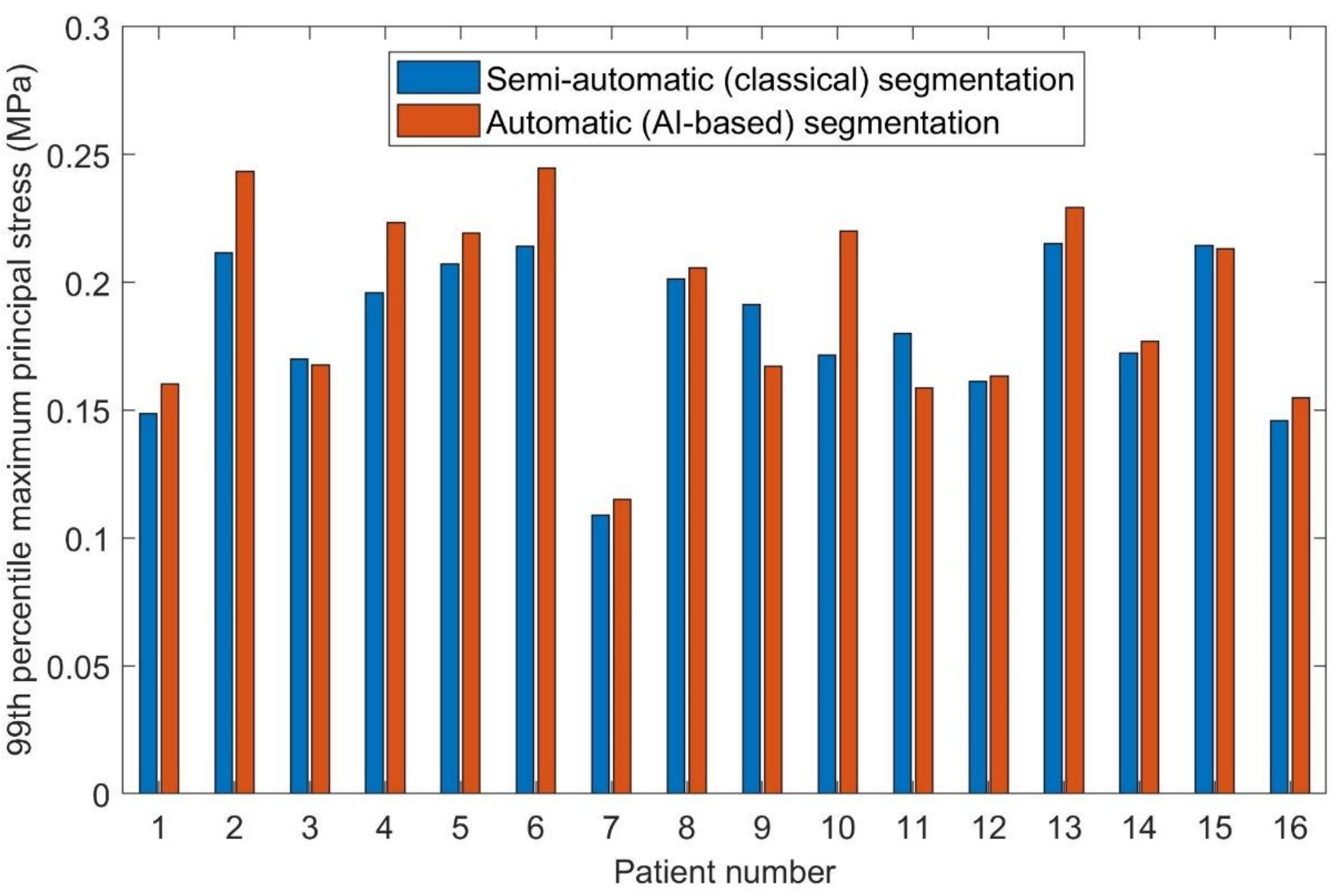

**Figure 8** Patient-by-patient comparison of the 99th percentile maximum principal stress obtained using AAA geometries segmented semi-automatically (classical method) and AAA geometries segmented automatically using an artificial intelligence (AI) algorithm.

# 4 Discussion

The stress distributions on AAA walls and their values, as seen in the maximum principal stress percentile plots in **Figure 7**, are very similar regardless of whether a fully automated AI-based or semi-automatic classical segmentation method was used. In fact, based on visual inspection, the difference between the percentile plots of the two methods in **Figure 7** is generally within the inter-analyst and intra-analyst variations in the maximum principal stress percentile plots computed from AAA models segmented by various analysts, as reported by (Hodge, Tan et al. 2023). Nevertheless, in 12 out of 16 cases, the peak and 99th percentile of maximum principal stress values obtained from AAA finite element models using geometries extracted using AI-based automatic segmentation are higher than those based on semi-automatic segmentation.

As we have previously explored the impact of the finite element mesh on stress results (Alkhatib, Bourantas et al. 2023, Alkhatib, Wittek et al. 2023), we are certain that our computed stresses are mesh-independent. This difference in maximum values of principal stress is due to variations in lumen segmentations. For 12 out of 16 cases, AI-based segmentation generated larger lumens than classical semi-automatic segmentations, and consequently larger surfaces on which pressure loads were applied in the finite element models, as discussed in more detail below.

### 4.1 Geometry and mesh

When comparing the number of elements in the ILT (**Table 3**), a notable difference emerged between semi-automated and automatic segmentations across most patients. Eight patients exhibited a relative absolute difference in the number of elements exceeding 10%. This difference is caused by slightly different smoothness of AAA wall surfaces, whose triangulations are used for initialization of automatic meshing. Our meshing method utilized the locations of nodes on the internal surface of the aneurysm wall as input for the external surface of the ILT, ensuring that surface nodes were shared and therefore that discretizations of the AAA wall and ILT were compatible (Alkhatib, Wittek et al. 2023).

The AI-based automatic segmentation generated a larger luminal surface compared to the luminal surface produced by the classical semi-automated segmentation (**Figure 9**). As the lumen surface is used to apply uniform pressure as loading in the finite element model, the differences in luminal surface areas directly translate to total loads and consequently to computed stress magnitudes. Consequently, in 12 out of 16 cases, higher values for the maximum principal stress were observed in finite element models extracted through the fully automated segmentation compared to their semi-automatically segmented counterparts.

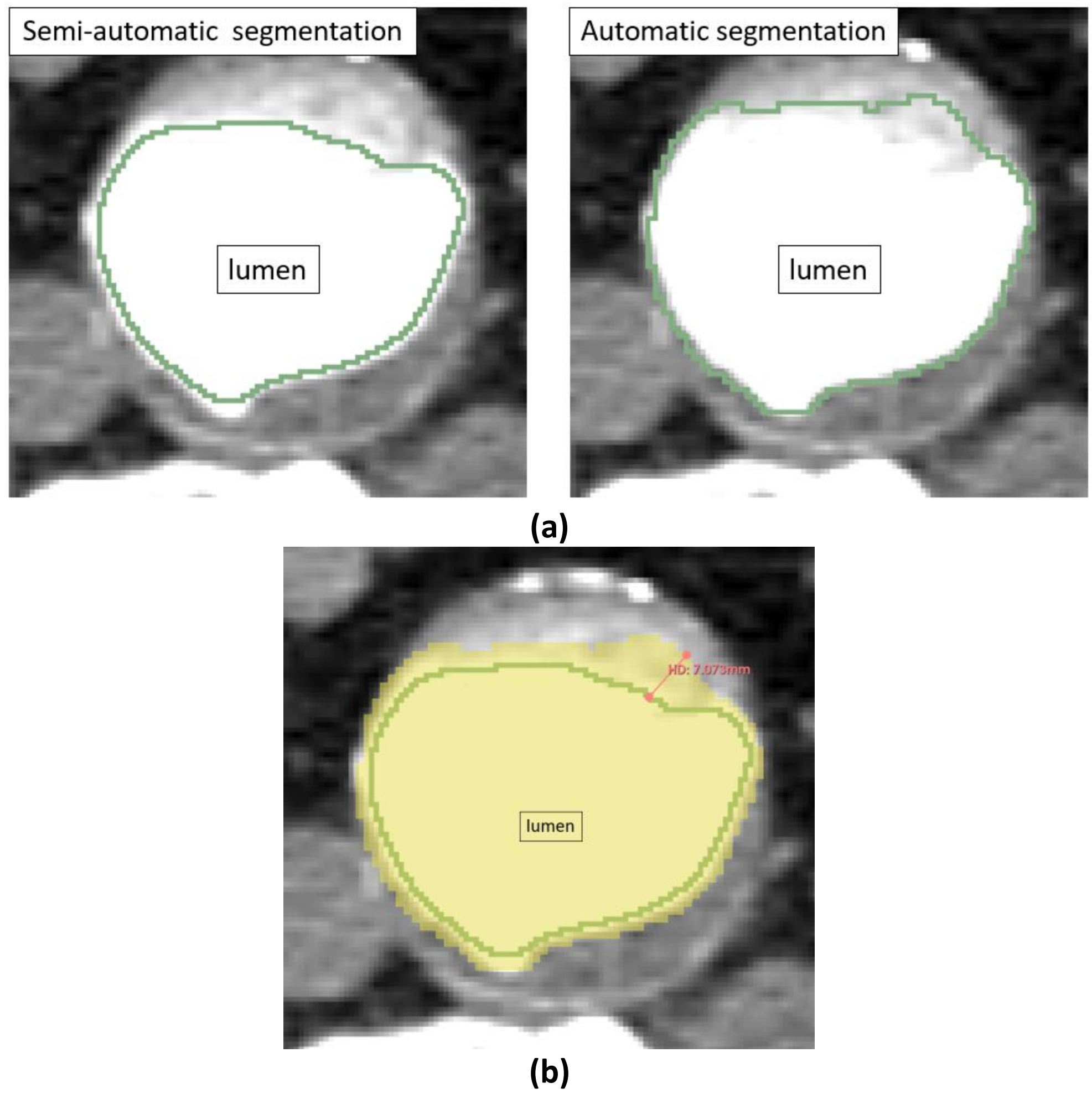

**Figure 9** Axial view of a slice from the computed tomography angiography (CTA) for an abdominal aortic aneurysm (AAA) (Patient 9) showing the lumen boundaries extracted using the threshold algorithm by semi-automatic segmentation where the threshold range was selected by an analyst and an automated segmentation where the threshold value was selected by an artificial intelligence (AI) algorithm. **(a)** Boundaries of the lumen are in green, and **(b)** the overlap of the two segmentations; yellow is the segmentation from the automatic segmentation and green boundary is the segmentation from the semi-automatic segmentation.

### 4.2 Lumen Segmentations Comparison

Due to uneven mixing of the contrast agent, determining the accurate lumen boundaries posed a challenge. In the absence of reliable ground truth, the question whether the semi-automatic or AI-based segmentation is "more correct" is difficult to answer. To gain additional insights into this important matter, we further expanded our investigation and calculated the Hausdorff distance (HD) (Rucklidge 1996, Garlapati, Mostayed et al. 2015) between the boundaries of the lumen segmented with both methods. HD was computed using the in-house MATLAB codes developed by the ISML team (Garlapati, Roy et al. 2014, Garlapati, Mostayed et al. 2015, Li, Miller et al. 2015). HD (**Equation 2**) is the maximum distance between the boundaries of the lumen segmentations.

$$\mathrm{HD}(\mathrm{X},\mathrm{Y}) = \max\left(\mathrm{h}(\mathrm{X}.\mathrm{Y}).\,\mathrm{h}(\mathrm{Y}.\mathrm{X})\right)\ , \qquad \textbf{(2)}$$

where X and Y represent the different segmentations

**Figure 10** shows the percentile plots of calculated Hausdorff distances within each slice of segmentation 2D axial view for the entire population of lumen boundaries segmented using both semi-automated and AI-based algorithm segmentations. As shown, the Hausdorff distances between contours are not negligible, and the larger lumens in the AI-segmented AAAs generally lead to higher maximum principal stresses. Nevertheless, statistical analysis conducted in Section 4.3 indicates that stress computation from AI-based segmentation is, for practical purposes, equivalent to that from classical segmentation.

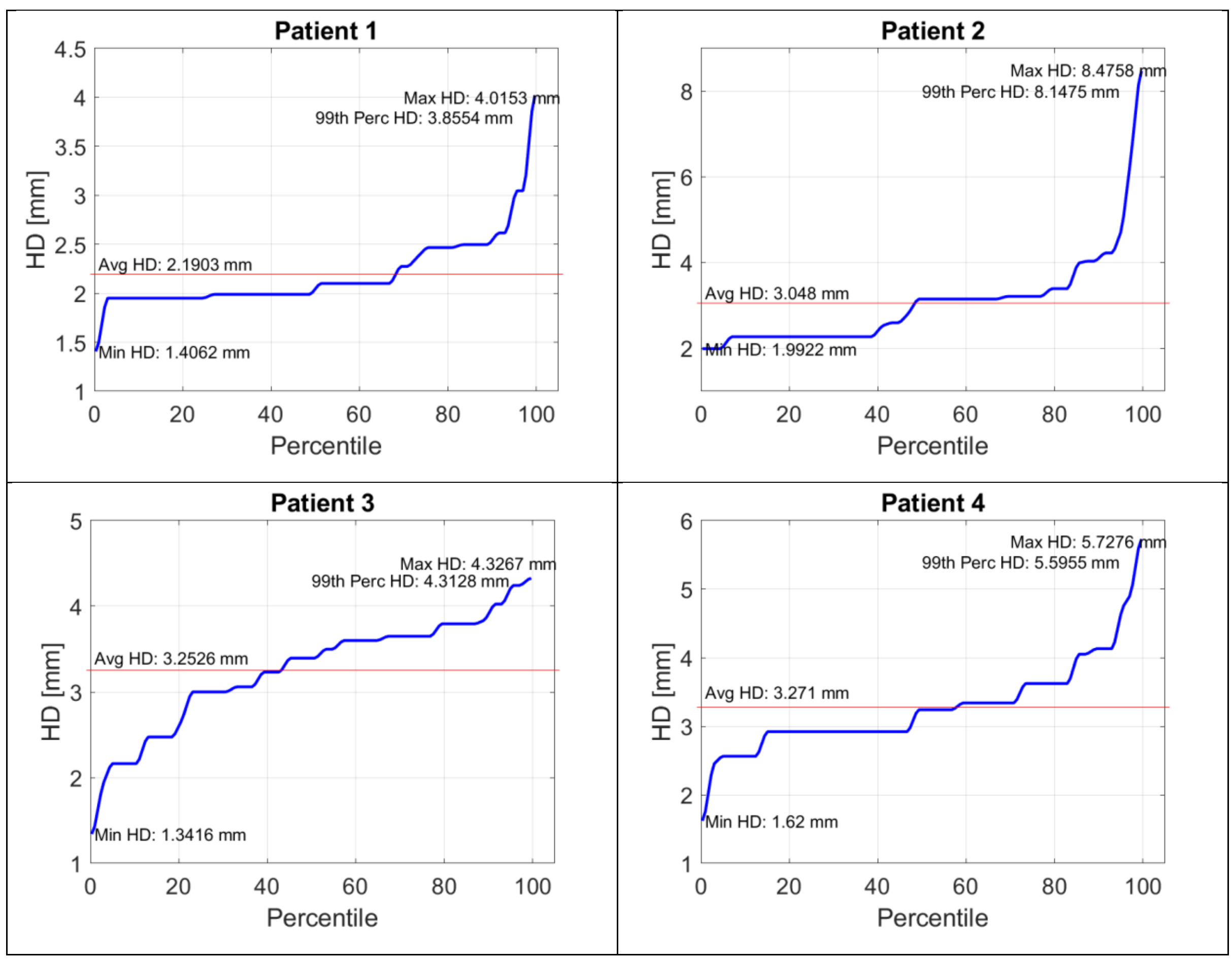

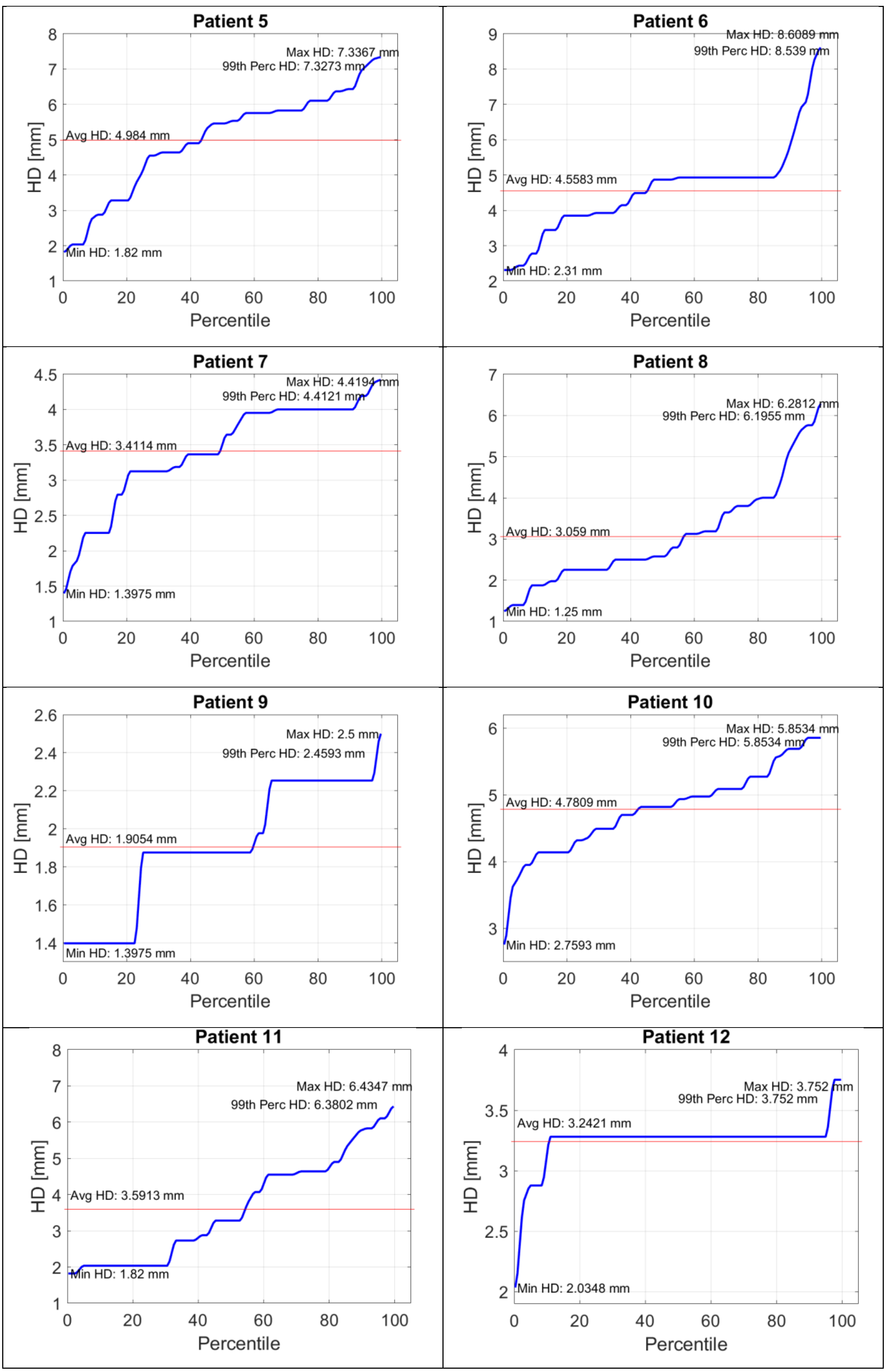

Patient 5
Max HD: 7.3367 mm
99th Perc HD: 7.3273 mm
Avg HD: 4.984 mm
Min HD: 1.82 mm
HD [mm]
Percentile
Patient 6
Max HD: 8.6089 mm
99th Perc HD: 8.539 mm
Avg HD: 4.5583 mm
Min HD: 2.31 mm
HD [mm]
Percentile
Patient 7
Max HD: 4.4194 mm
99th Perc HD: 4.4121 mm
Avg HD: 3.4114 mm
Min HD: 1.3975 mm
HD [mm]
Percentile
Patient 8
Max HD: 6.2812 mm
99th Perc HD: 6.1955 mm
Avg HD: 3.059 mm
Min HD: 1.25 mm
HD [mm]
Percentile
Patient 9
Max HD: 2.5 mm
99th Perc HD: 2.4593 mm
Avg HD: 1.9054 mm
Min HD: 1.3975 mm
HD [mm]
Percentile
Patient 10
Max HD: 5.8534 mm
99th Perc HD: 5.8534 mm
Avg HD: 4.7809 mm
Min HD: 2.7593 mm
HD [mm]
Percentile
Patient 11
Max HD: 6.4347 mm
99th Perc HD: 6.3802 mm
Avg HD: 3.5913 mm
Min HD: 1.82 mm
HD [mm]
Percentile
Patient 12
Max HD: 3.752 mm
99th Perc HD: 3.752 mm
Avg HD: 3.2421 mm
Min HD: 2.0348 mm
HD [mm]
Percentile

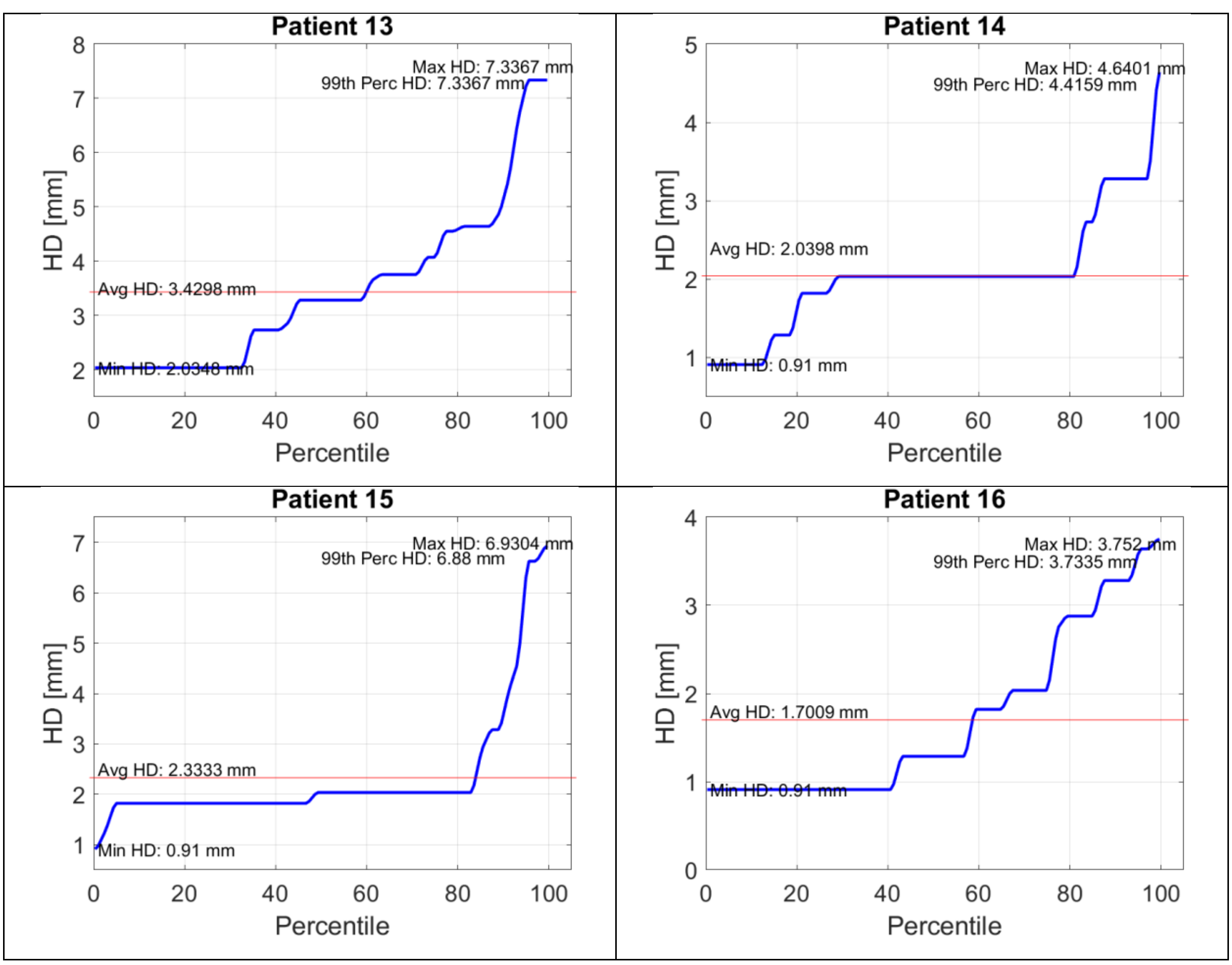

**Figure 10** Hausdorff Distance (HD) in mm between the segmented lumen boundaries in both segmentation methods, semi-automatic and automatic segmentations, against the percentile of edges in the axial image slices. The red dashed line indicates the average value of the HD distance.

## 4.3 Statistical analysis

We applied Student's t-test to test the null hypothesis that the difference in means of the 99th percentile maximum principal stress from the AAA finite element models using AI-based automatic segmentation and from the models using semi-automatic segmentation is zero. The test yielded a p-value of 0.438 which implies that the null hypothesis cannot be rejected at the 0.05 level typically used in the literature. The high p-value of 0.438 indicates that this hypothesis cannot be rejected at any reasonable level of significance, thereby confirming our conclusion that stress computation from AI-based segmentation is, for practical purposes, equivalent to classical segmentation.

This result is consistent with the analysis using box plots (whisker plots). As shown in **Figure 11**, the difference between the median 99th percentile maximum principal stress (0.1913 MPa) from the AAA finite element models using AI-based automatic segmentation and the median (0.1856 MPa) from models using semi-automatic segmentation is much smaller than the interquartile range for 99th percentile maximum principal stress values obtained with semi-automatic segmentation. In addition, the difference between these two medians (0.0057 MPa) is much smaller than the inter-analyst (0.1252 MPa) and intra-analyst (0.1253 MPa) differences in the 99th percentile maximum principal stress computed from AAA models segmented by various analysts, as reported in the literature (Hodge, Tan et al. 2023).

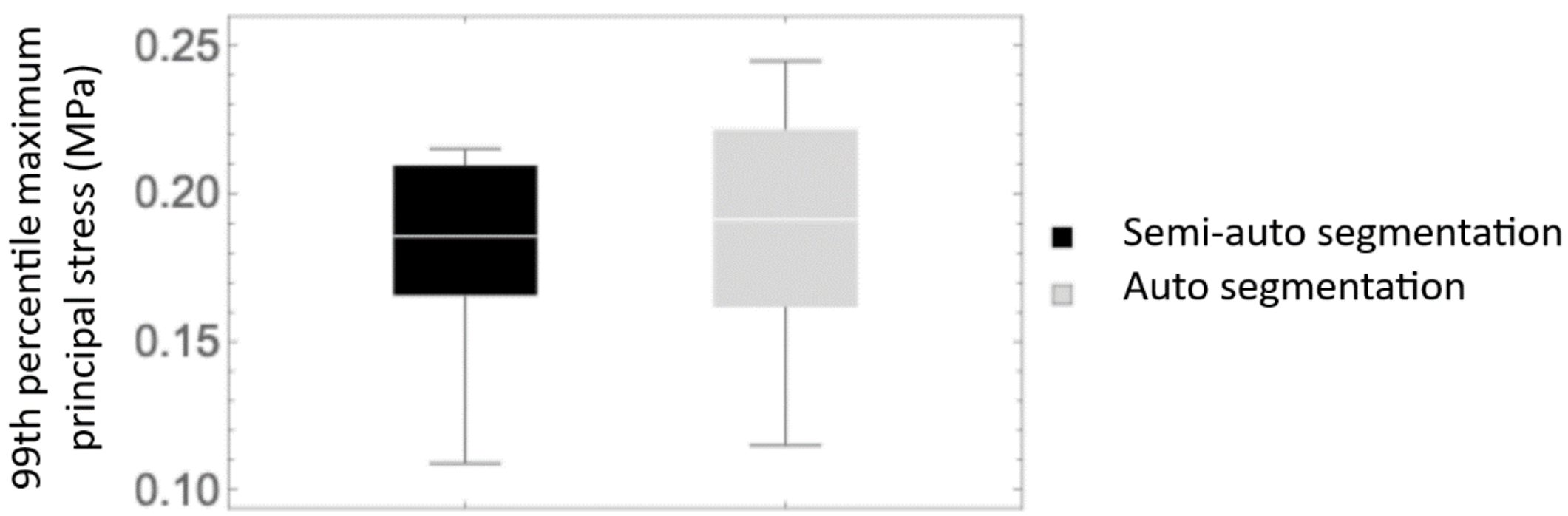

**Figure 11** Comparison between the 99th percentile maximum principal stress obtained using AAA geometries segmented semi-automatically (Classical method) and AAA geometries segmented automatically using artificial intelligence algorithm (AI).

# 5 Conclusions

We have developed a fully automated pipeline that begins with the patient's AAA CTA images and concludes with the calculation of stress in the AAA wall. This automated process integrates three sequential, fully automated subprocesses: (1) AI-based segmentation using the PRAEVAorta2® web portal, (2) automatic post-processing using our in-house MATLAB code, and (3) Computation of stress in the AAA wall using BioPARR. The entire pipeline is fully automated, with the output of each subprocess serving as the input for the next, and there is no manual intervention at any stage.

The findings of this study demonstrate a good correspondence between the stress calculations for abdominal aortic aneurysms (AAA) geometries extracted using a classical semi-automated segmentation and those generated through a fully automated segmentation employing an artificial intelligence (AI) algorithm. The peak and 99th percentile of maximum principal stress generally exhibited slightly higher values in the finite element models extracted through automatic segmentation compared to semi-automatic segmentation. However, no differences between the overall distributions of the maximum principal stress could be detected through qualitative analysis using visual inspection. Furthermore, our statistical analysis indicates that the differences in AAA wall stress obtained using the two segmentation methods are not statistically significant and fall within the typical inter-analyst and intra-analyst variability, as reported in (Hodge, Tan et al. 2023). These results may be interpreted as confirming the accuracy of a fully automated AAA stress computation pipeline using AI-based automatic segmentation.

This consistent difference in calculated maximum stress values was attributed to the AI algorithm's segmented lumens consistently exhibiting a larger surface area compared to semi-automatically segmented ones, resulting in larger total pressure load on the finite element model of AAA. This difference points to the importance of determining the lumen boundaries, a problem whose difficulty partly stems from uneven mixing of a contrast agent used during imaging, and that is frequently underestimated in the literature. Moreover, our semi-automated lumen segmentation approach, based on thresholding, does not account for partial volume effects at the lumen boundary. Despite an analyst's effort to correct this manually, the lumen volume may still be underestimated. It also points to the need to rely on some normalized measure of stress, perhaps scaled by an "average stress", rather than maximum (or 99th percentile) values.

In this paper, we demonstrated a fully automated pipeline — from CT image of abdomen containing an aortic aneurysm to stress computation — relying on AI-based segmentation and a novel segmentation post-processing. Our fully automated pipeline integrates the Nurea commercial software for AI-based segmentation with the well-established open-source BioPARR software package for biomechanical computations (https://bioparr.mech.uwa.edu.au/), through our novel and fully automated post-processing of AI-based segmentations. The effectiveness and appropriateness of our methods were demonstrated on patients from three hospitals (in Australia, Austria and Belgium) and images obtained using three different scanners and scanning protocols, confirming the robustness and potential clinical applicability of our work.

## Acknowledgments

This work was supported by the Australian National Health and Medical Research Council NHMRC Ideas grant no. APP2001689. The contributions of Christopher Wood and Jane Polce, radiology technicians at Medical Imaging Department, Fiona Stanley Hospital (Western Australia, Australia) in patient image acquisition at Fiona Stanley Hospital are gratefully acknowledged.